\documentclass[12pt]{article} 
\usepackage{amsmath}
\usepackage{graphicx}
\usepackage{enumerate}
\usepackage{natbib}
\usepackage{multirow}
\usepackage{longtable}
\usepackage{booktabs}
\usepackage{amssymb}
\usepackage{bm}
\usepackage{mathrsfs}
\usepackage{amsthm}
\usepackage{amsfonts}
\usepackage{subfigure}
\usepackage{floatrow}
\usepackage{epsfig,amssymb,latexsym,verbatim}
\usepackage{epstopdf}
\usepackage{graphics}
\usepackage[english]{babel}
\usepackage[figuresright]{rotating}
\usepackage[dvipsnames]{xcolor}
\usepackage{color}
\usepackage{hyperref}

\usepackage{xr}
\allowdisplaybreaks
\newtheorem{theorem}{Theorem}

\newtheorem{corollary}{Corollary}
\numberwithin{equation}{section}
\numberwithin{lemma}{section}
\numberwithin{theorem}{section}
\numberwithin{prop}{section}
\numberwithin{corollary}{section}

\def\P{\mathbb{P}}
\def\E{\mathbb{E}}
\def\Var{\mbox{Var}}

\def\o{{\scriptstyle{\mathcal{O}}}}
\def\O{\mathcal{O}}

\newcommand{\blind}{0}

\begin{document}

\def\spacingset#1{\renewcommand{\baselinestretch}%
{#1}\small\normalsize} \spacingset{1}

\date{}
\if0\blind
{
  \title{\bf From sparse to dense functional time series: phase transitions of detecting structural breaks and beyond}
  \author{Leheng Cai\footnotemark[1] $\,$ and$\,$  Qirui Hu\footnotemark[1] \footnotemark[2] \hspace{.2cm}}
  \maketitle
  \renewcommand{\thefootnote}{\fnsymbol{footnote}}

  \footnotetext[1]{Department of Statistics and Data Science,Tsinghua University}
           
		\footnotetext[2]{Corresponding author: hqr20@mails.tsinghua.edu.cn}
} 

\if1\blind
{
  \bigskip
  \bigskip
  \bigskip
  \begin{center}
    { \bf From sparse to dense functional time series: phase transitions of detecting structural breaks and beyond}
\end{center}
  \medskip
} \fi

\bigskip

\maketitle
	
	\textbf{Abstract:}   We develop a novel methodology for detecting abrupt break points in  mean functions of functional time series, adaptable to arbitrary sampling schemes.  By employing B-spline smoothing, we introduce $\mathcal L_{\infty}$ and $\mathcal L_2$ test statistics statistics based on a smoothed cumulative summation (CUMSUM) process, and derive the corresponding asymptotic distributions  under the null and local alternative hypothesis, as well as the phase transition boundary from sparse to dense.   We further establish the convergence rate of the proposed break point estimators and conduct statistical inference on the jump magnitude  based on the estimated break point, also applicable across sparsely, semi-densely, and densely,   observed random functions. Extensive numerical experiments validate the effectiveness of the proposed   procedures. To illustrate the practical relevance, we  apply the developed methods to analyze electricity price data and   temperature data.


	\maketitle
	
\newpage
\spacingset{1.9} 
\section{Introduction}
\par Functional time series analysis has emerged as a prominent field in modern statistics, with models and methodologies being applied across various disciplines, such as  neuroscience, sociology, economics, and environmental science. Standard references include \cite{bosq2000linear} and \cite{horvath2012inference}. 

\begin{figure}[H]
    \centering    \includegraphics[width=0.7\linewidth]{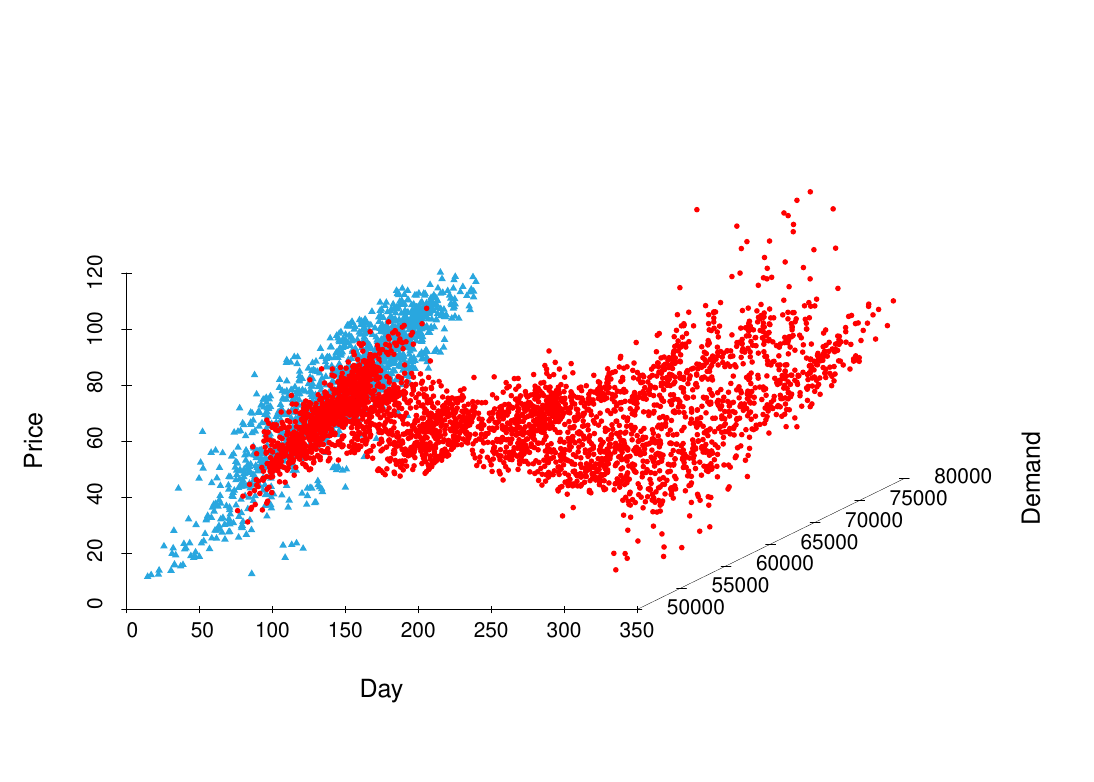}
    \caption{Daily electricity price-demand data of German in 2011. Before March 15, triangle markers; after March 15, circle markers. }
    \label{fig:3dplot}
\end{figure}

Stationarity is a common assumption in the existing literature on functional time series; see \cite{10.1214/09-AOS768}, \cite{horvath2013estimation} and \cite{HuStatisticalIF} However, some real-world functional data collected in timely order exhibit structural breaks, leading to a violation of the stationarity assumption.  
An example of such data 
is illustrated in Figure \ref{fig:3dplot}, which displays the daily electricity price-demand data of German in 2011. Due to  the Fukushima nuclear
leak in Japan,  the German government immediately enacted a new electricity supply policy on March 15, leading to a abrupt  shift in the electricity price-demand structure. A detailed analysis of this dataset is referred to section \ref{sec:realdata}.

 \par 
  The following model summarizes the key features of such  observed data $\left\{(X_{ij},Y_{ij})\right\}_{i,j=1}^{n,N_i}$,
\begin{align}\label{model1}
		Y_{ij} = m_i(X_{ij}) + \xi_i(X_{ij}) + \sigma(X_{ij})\varepsilon_{ij}, \quad i=1,\cdots,n, \, j=1,\cdots, N_i,
	\end{align}
    in which the deterministic functions
    $\{m_i(\cdot)\}_{i=1}^{n}$ denotes the mean functions. The  functions $\left\{\xi_i(\cdot)\right\}_{i=1}^{n}$ form a stationary sequence of square-integrable   stochastic processes with mean zero over $[0,1]$, which can be viewed as $\mathcal L^2[0,1]$-valued random elements. 
   Technically, one has $\xi_i(x) \in \mathcal{L}^2[0,1]$ almost surely, satisfying $\mathbb{E}\int_0^1\xi_i^2(x)dx<\infty$. The discrete girds 
    $\{X_{ij}\}_{i,j=1}^{n,N_i}$ denotes the locations where the random trajectories $\{\xi_i(\cdot)\}_{i=1}^{n}$ are recorded, and are i.i.d. copies of $X$ with density $f(\cdot)$. 
    The measurement errors $\varepsilon_{ij}$'s are i.i.d. random variables such that  $\E\left(\varepsilon_{ij}\right) = 0$ and $\Var\left(\varepsilon_{ij}\right) = 1$, with variance function  $\sigma^2(\cdot)$. 
Besides, $\left\{N_i\right\}_{i=1}^{n}$ is an array of i.i.d. random variables, taking values in $\mathbb Z_+$, and $\P(N_i >1)>0$, and the sets $\{\xi_{i}(\cdot)\}_{i=1 }^{n }$, $\{X_{ij}\}_{i=1,j=1}^{n,N_i}$, and $\{\varepsilon_{ij}\}_{i=1,j=1}^{n,N_i}$  are mutually independent.  While $\{N_i\}_{i=1}^{n}$ is assumed to be an i.i.d. array, we do not require $\mathbb{E}(N_i)$ is bounded, but it can grow to infinity  as $n$ increases. 

\par A  fundamental and important task is to test for the presence of structural breaks in  mean functions over time, and further  estimate the locations of the break if exists. 
  To address this critical issue, we first consider the following  break point testing problem in mean functions, where the null and alternative hypotheses are defined as:
\begin{align}\label{EQ:hypothesis}
    H_0: m_1 = \cdots = m_n \quad \mbox{versus} \quad H_1: m_1 = \cdots = m_{k_0} \neq m_{k_0 + 1}  \cdots = m_n.
\end{align}
Here, $k_0$ is the break point such that $k_0/n \in (\epsilon, 1-\epsilon)$ for some fixed $\epsilon \in (0, 1/2)$. 

The initial approach relies on   projection methods to estimate   functional principal component (FPC) scores and then transforms the hypothesis in (\ref{EQ:hypothesis}) into a finite-dimensional testing problem, as discussed in \cite{aue2009estimation} and \cite{berkes2009detecting}. These methods require   consistent estimators for FPC scores and appropriate truncation for the number of FPCs. Also, the projection approach cannot handle the cases when the jump (defined below) is orthogonal to the FPCs. To proceed,  \cite{aue2018detecting} proposed a CUMSUM test statistic based on the $\mathcal{L}_2$-norm, employing the weak invariance principle in Hilbert space $\mathcal{L}^2[0,1]$ to avoid dimension reduction. This pioneering research has been widely applied to detect structural breaks in other complex types of functional data, including high-dimensional, spatially correlated, and long-range dependent cases, as illustrated in \cite{li2023detection}, \cite{wang2023asynchronous}, and \cite{baek2024test} respectively.  Power enhanced methods have also been considered, see \cite{jiao2023enhanced}. Additionally,  generalized CUMSUM processes have been adapted for estimating multiple break points, as discussed in \cite{rice2022consistency}, \cite{bastian2023multiple}, and \cite{harris2022scalable}. However,   the aforementioned literature on testing break points  focuses on ideal scenarios where trajectories can be fully collected, whereas in practice, recordings are always  occur only at regular or random locations. Recently, \cite{hu2024change} addressed the break point analysis of functional variance functions based on data contaminated by measurement errors, by applying B-spline pre-smoothing and kernel methods to asymptotically ignore the noise effect in a dense setting. Moreover, \cite{BHW24} developed testing procedures for locally stationary functional time series without a pre-smoothing step, accommodating both dense and semi-dense scenarios. These studies, however, only consider regularly recorded data and assume  that the sampling frequency    be of a comparable order to the sample size $n$.

\par On the other hand,  functional data analysis from sparse to dense and related phase transitions are attracting increasing attention. \citet{ZW16} conducted an in-depth analysis of mean and covariance functions, meticulously investigating the convergence rates and point-wise asymptotic normality transitioning from sparse to dense functional data. They introduced subcategories such as sparse, semi-dense and  dense, based on the achievement of the $n^{-1/2}$ rate and the neglect of asymptotic bias. \citet{sharghi2021mean} explored the estimation of mean derivatives in longitudinal and functional data across different sampling schemes.    Besides,  \citet{tonycai2011} and \citet{berger2023dense} investigated the optimal rates for estimating the mean function in the $\mathcal{L}_2$ and $\mathcal{L}_{\infty}$ norms, respectively. More recently, \citet{Zhang2021UnifiedPC} delved into a unified principal component analysis for sparse and dense functional data, especially under conditions of spatial dependency. \citet{guo2023sparse} achieved significant breakthroughs in non-asymptotic mean and covariance estimation within high-dimensional functional time series. From the statistics inference perspective, \cite{cai2024sparse} constructed simultaneous confidence bands for the mean function transitioning from sparse to dense, utilizing B-spline and other orthogonal series estimator. In the realm of break points analysis in functional data, to the best of our knowledge, only \cite{madrid2022change} and \cite{xue2024change} have considered the estimation of break points in mean function and covariance structures, respectively, under the assumption of existing breaks. However, for testing whether there is a structural break in the mean function,  existing work   typically assume specific sampling schemes and develop corresponding procedures  based on their restrictive assumptions, thus no general approaches without sampling restrictions have been studied yet. 

%


In this paper, one innovation that should be emphasized is that our developed approach provides a unified framework for testing the break point in functional time series, valid under a general dependence structure and allowing for arbitrary sampling designs. By employing B-spline estimators for the partial and global mean functions, we construct a smoothed CUMSUM process. Beyond the conventional $\mathcal{L}_2$-norm based test statistic, we also introduce an $\mathcal{L}_\infty$-norm based test statistic. Although the theoretical properties of $\mathcal{L}_2$ statistics have been widely studied, analyzing $\mathcal{L}_\infty$ statistics is more challenging due to the absence of an inner product in the space $\mathcal C[0,1]$. Each of these two statistics has its own advantages depending on the features of the jump. Our work rigorously derives the general asymptotic distributions of the proposed test statistics under the null and local alternatives, and further establishes the phase transition boundaries for sparsely, semi-densely, and densely observed data. Since the smoothed CUMSUM process we constructed might not be tight, leading  to an  additional gap to weak convergence results, we alternatively provide   Gaussian approximation varying with time $n$. The derived theory is consistent with the specific cases, such as those discussed in dense or semi-dense scenarios of \cite{BHW24} and fully observed scenarios of \cite{aue2018detecting}. Under certain conditions, the approximating Gaussian fields will reduce to the ones in both papers. Recent results on sequential Gaussian approximation under physical dependence from \cite{mies2023sequential}, the L\'{e}vy concentration inequality in \cite{chernozhukov2015comparison}, and properties of splines from \cite{de1978practical} and \cite{schumaker2007spline} are interwoven seamlessly to achieve satisfactory theoretical results.

Additionally, we develop two estimators for break points based on the aforementioned $\mathcal{L}_{\infty}$ and $\mathcal{L}_2$ statistics, and investigate the corresponding convergence rates. Moreover, we are interested in the simultaneous inference of the jump magnitude,  defined by $\Delta(\cdot) = m_{k_0+1}(\cdot) - m_{k_0}(\cdot)$. A two-step unified simultaneous confidence band (SCB) for the jump magnitude  is constructed. SCBs are powerful tools for quantifying the variability of complex functions
and making  statistical inference, thus it is highly desirable to construct SCBs for the jump magnitude backed by rigorous theory. So far, we have provided a comprehensive   methodology for the hypothesis testing problem~\eqref{EQ:hypothesis}, along with subsequent break point location and jump inference, from sparse to dense functional time series.

\par This paper is organized as follows: Section \ref{sec:method} introduces the   construction of the smoothed CUMSUM process, along with the proposed test statistics,   break point estimators, and jump magnitude estimators. Section \ref{sec:main results} establishes    corresponding asymptotic properties. Section \ref{sec:implement} provides detailed descriptions of the implementation. Simulation results and data applications are presented in Sections \ref{sec:simulation} and \ref{sec:realdata}, respectively. Technical proofs of theoretical  results are included in the supplemental materials.

\section{Methodology}\label{sec:method}

\par To describe the temporal dependence, following \cite{zhou2023statistical} we specify $\xi_i(\cdot) = H(\cdot,\mathcal F_i)$, where $H: [0,1] \times \mathcal S^{\mathbb N} \to \mathbb R$ is a filter, $\mathcal F = (\cdots, e_{i-1}, e_i)$, and $\{e_i\}_{i \in \mathbb Z}$ is a sequence of i.i.d. random elements within some measurable space $\mathcal S$.  Then, the physical dependence measure   of $\{\xi_i(\cdot)\}_{i \in \mathbb Z}$ is defined by: $ 
    \delta_H(i,r) = \sup_{x \in [0,1]}\left\|H(x, \mathcal F_i) - H(x, \mathcal F^\ast_i)\right\|_r$, 
where $\mathcal F^\ast_i = (\cdots, e_{-1}, e_0^\ast, e_1, \cdots, e_i)$, and $e_0^\ast$ is a copy of $e_0$ and independent of $\{e_i\}_{i \in \mathbb Z}$. The long-run covariance function of $\{\xi_i(\cdot)\}_{i=1}^{n}$ is expressed as:
	\begin{align*}
	G(x,x^\prime) = \sum_{h=-\infty}^{\infty} \mbox{Cov}\left(H(x, \mathcal F_h), H(x^\prime, \mathcal F_0)\right).
\end{align*}
Common stationary functional time series models include Functional AR and Functional MA models. The details of their physical dependence measures, as well as examples of other nonlinear functional time series models, are referred to \cite{zhou2023statistical} and \cite{dette2021confidence}. 


\par In the following, denote $\eta_i(\cdot)=m(\cdot)+\xi_i(\cdot)$ for $i=1,\ldots,n$.
As pointed in \cite{aue2018detecting}, if one fully observes the random functions $\{\eta_i(\cdot)\}_{i=1}^{n}$, the test statistic  could be  the squared $\mathcal{L}_2$-norm of a CUMSUM process, i.e.,
\begin{align}\label{EQ:Snfull}
    \widetilde{S}_n = \max_{1\leq k\leq n}\left\|\frac{1}{\sqrt{n}}\left(\sum_{i = 1}^k \eta_i(x) - \frac{k}{n}\sum_{i=1}^n \eta_i(x)\right)\right\|_{\mathcal{L}_2}^2.
\end{align}
Under some regular conditions, $\widetilde{S}_n$  weakly converges to some functional of a Gaussian field with covariance function $ \left(\min\{t,t^\prime\}-tt^\prime \right)G(x,x^{\prime})$. For sufficient dense collected data, one may recover the trajectories  $\left\{\eta_i(\cdot)\right\}$ by some non-parametric smoothers, denoted by $\left\{\widehat{\eta}_i(\cdot)\right\}_{i=1}^{n}$, and then considers the squared $\mathcal{L}_2$-norm of  a plug-in CUMSUM process,  
\begin{align}\label{EQ:dense}
    \widehat{S}_n = \max_{1\leq k\leq n}\left\|\frac{1}{\sqrt{n}}\left(\sum_{i = 1}^k \widehat{\eta}_i(x) - \frac{k}{n}\sum_{i=1}^n \widehat{\eta}_i(x)\right)\right\|_{\mathcal{L}_2}^2.
\end{align}
By proving the oracle efficiency property  as described in  \cite{cao2012simultaneous} 
or \cite{caiHu}, $\widehat{S}_n$ is asymptotically equivalent to $\widetilde{S}_n$. When the recorded location $X_{ij}$ are fixed and regular, i.e., $X_{ij} = j/N$ for all $i=1,\ldots,n$, \cite{BHW24} proposed the following statistics without pre-smoothing and allows the data   observed semi-densely,
\begin{align}\label{EQ:Snsemidense}
   \check{S}_n = \max_{1\leq k \leq n}\frac{1}{\sqrt{n} N}\sum_{j = 1}^{N}\left(\sum_{i = 1}^{k}Y_{ij} -\frac{k}{n}\sum_{i = 1}^{n}Y_{ij}\right)^2. 
\end{align}
As $n,N\to\infty$, the asymptotically distribution of $\check{S}_n$ is related to the long-run covariance function $G(x,x^{\prime})$ of the processes $\{\xi_i(\cdot)\}_{i=1}^{n}$ and the variance function $\sigma^2(x)$ of the  errors $\varepsilon_{ij}$'s.

For practically collected data in (\ref{model1}), it is hard to approach the conditions of oracle efficiency for $\widehat{S}_n$ in  (\ref{EQ:dense}) when the sampling scheme is not dense enough. Beside,  the test statistic $\check{S}_n$ in (\ref{EQ:Snsemidense}) can not accommodate the settings of irregular design, also their proposed  bootstrap method   would fail. Therefore,  it is strongly motivated to develop a  unified theory for the break point testing problem in (\ref{EQ:hypothesis}), especially for sparse functional times series.

\par Before introducing our newly proposed statistics,  B-spline functions are described below. Let $\{t_l\}_{l=0}^{J_n+1}$ be a sequence of equally-spaced points, where $t_l = l / (J_n+1)$ for $0 \leq l \leq J_n+1$, which divide $[0,1]$ into $J_n+1$ equal sub-intervals, denoted as $I_l = [t_l, t_{l+1})$ for $l = 0, \ldots, J_n-1$, and $I_{J_n} = [t_{J_n}, 1]$. Let $\mathcal{S}^{p}_{J_n} = \mathcal{S}^{p}_{J_n}[0,1]$ be the polynomial spline space of order $p$ over $
\left\{I_l\right\}_{l=0}^{J_n}$, consisting of all functions that are $(p-2)$ times continuously differentiable on $[0,1]$ and are polynomials of degree $(p-1)$ within the sub-intervals $I_l$, $l = 0,\ldots,J_n$. Let $\bm B(x)=\{B_{l,p}(x),\ldots,B_{J_n+p,p}(x)\}^\top$ be the $p$-th order B-spline basis functions of $\mathcal{S}_{J_n}^{p}$, then $\mathcal{S}_{J_n}^{ p}=\left\{
 \sum_{l =1}^{J_n+p}\lambda _{l}B_{l,p}(x
): \lambda _{l}\in \mathbb{R} \right\} $.

\par We construct a smoothed CUMSUM process that can  adapt to functional time series from sparse to dense by pooling the data points to estimate the partial mean and global mean function separately. For some fixed $\epsilon\in(0,1/2)$ and any $t \in (\epsilon, 1- \epsilon)$,  the estimation of partial mean function and the global mean function   is given by 
\begin{align}
&\label{unweighted:mhat_t}
    \widehat{m}(t,x) = \mathop{\arg\min}
\limits_{g(\cdot)\in \mathcal{S}_{J_n}^{p}}
\frac{1}{[nt]}\sum_{i=1}^{[nt]} \frac{1}{N_i}\sum_{j=1}^{N_i}  \left\{Y_{ij}-g
(X_{ij})\right\}^2,\\
    &\label{unweighted:mhat}
\widehat{m}(x)=\mathop{\arg\min}
\limits_{g(\cdot)\in \mathcal{S}_{J_n}^{p}}
\frac{1}{n}\sum_{i=1}^n \frac{1}{N_i}\sum_{j=1}^{N_i}  \left\{Y_{ij}-g
(X_{ij})\right\}^2.
\end{align}
The solution to (\ref{unweighted:mhat_t}) and (\ref{unweighted:mhat}) are $\widehat m(t,x)=\bm B^\top (x)\widehat{\bm\theta}_t$ and $\widehat m(x)=\bm B^\top (x)\widehat{\bm\theta}$  respectively, where
\begin{align*}
		&\widehat{\bm\theta}_t =\widehat{ \mathbf{V}}_t^{-1}\left\{\frac{1}{[nt]}\sum_{i=1}^{[nt]} \frac{1}{N_i}\sum_{j=1}^{N_i} \bm B(X_{ij})Y_{ij}\right\},\quad
		\widehat{\mathbf{V}}_t= \frac{1}{[nt]}\sum_{i=1}^{[nt]} \frac{1}{N_i}\sum_{j=1}^{N_i} \bm B(X_{ij}) \bm B^\top (X_{ij}).\\
  &\widehat{\bm\theta} =\widehat{ \mathbf{V}}^{-1}\left\{\frac{1}{n}\sum_{i=1}^{n} \frac{1}{N_i}\sum_{j=1}^{N_i} \bm B(X_{ij})Y_{ij}\right\},\quad
		\widehat{\mathbf{V}}= \frac{1}{n}\sum_{i=1}^{n} \frac{1}{N_i}\sum_{j=1}^{N_i} \bm B(X_{ij}) \bm B^\top (X_{ij}).
\end{align*} 
Methodologically, different from \cite{madrid2022change} that assumes that the number of observations is the same across curves and adopts  Nadaraya-Watson estimators, we consider more computationally efficient B-spline estimators, which  assign   different weights $N_i^{-1}$ to the data points on $i$-th trajectory to accommodate different sampling frequency across different trajectories. Similar   weighted smoothing estimators can be found in \cite{10.1214/10-AOS813} and  \cite{tonycai2011}. 
Theoretically, inspired by recent results of sequential Gaussian approximation  in high dimensions in \cite{mies2023sequential},  the definition in (\ref{unweighted:mhat_t}) allows 
$\widehat{\bm\theta}_t$  to be expressed as the partial mean of a vector-valued stationary time series,  multiplied on the left by the inverse of an  inner product matrix  under the null  in (\ref{EQ:hypothesis}). This fact is crucial for constructing   Gaussian approximation of the partial mean and global mean estimators.

Then, we adopt the partial mean and   global mean to form a smoothed CUMSUM process, and further develop the $\mathcal L_{\infty}$ and  $\mathcal L_{2}$ statistics, $T_n$ and $S_n$, respectively, to address different types of jump functions.
\begin{align}
   &T_n =  \sup_{\epsilon\leq t\leq 1-\epsilon}\left\|   \frac{  [nt]   \widehat{m}(t,\cdot)/ {\sqrt{n}}  -    {\sqrt{n}t} \widehat{m}(\cdot)  }{\sqrt{ \bm B^\top(\cdot)\mathbf\Sigma\bm B(\cdot)}}\right\|_{\infty},\label{EQ:Tn}\\
   & S_n = \sup_{\epsilon\leq t\leq 1-\epsilon} \left\|   \frac{  [nt]   \widehat{m}(t,\cdot)/ {\sqrt{n}}  -    {\sqrt{n}t} \widehat{m}(\cdot)  }{\sqrt{ \bm B^\top(\cdot)\mathbf\Sigma\bm B(\cdot)}}\right\|_{\mathcal{L}_2}^2,\label{EQ:Sn}
\end{align}
where   $\mathbf \Sigma =\mathbf \Sigma_1+\mathbf \Sigma_2$,  $\mathbf{V} = \E \bm B(X)\bm B^\top(X)$, and
\begin{equation}\label{Sigma}
    \begin{aligned}
    &\mathbf \Sigma_1=
    \mathbf{V}^{-1} \E\left\{ (N-1)/N\right\}\E \bm B(X)\bm B^\top(X^\prime)G(X,X^\prime)\mathbf{V}^{-1},\\&
    \mathbf \Sigma_2 = \mathbf{V}^{-1} \E (N^{-1})\E \bm B(X)\bm B^\top(X)\left\{G(X,X)+\sigma^2(X)\right\}\mathbf{V}^{-1}.
    \end{aligned}
\end{equation}
Here, $X^\prime$ is an i.i.d. copy of $X$.
 It is worth noticing that the range of supreme with respect  to $t$ can be relaxed to $[\epsilon,1]$, while the lower bound needs to be strictly positive, since one needs the inverse of $\widehat{\mathbf V}_t$  to be well defined.
Observe that the smoothed CUMSUM process,  as a function of $t$,  tends to grow large in terms of either $\mathcal L_\infty$ or $\mathcal L_2$ norm at the rescaled break point $\tau_0=\lim_{n\to\infty}k_0/n$. Hence, one rejects the null when the value of $T_n$ or $S_n$ is large.  Notice that the numerators in $T_n$ and $S_n$ form a CUMSUM process, which can be approximated by  (\ref{EQ:Snfull}) and (\ref{EQ:dense}) in the  completely and densely observed functional time series setting. However, an additional unknown term appears in the denominators and serves as a normalizer, preventing the statistics from diverging, particularly in sparse data scenarios. A detailed discussion on $\bm{B}^\top(\cdot)\mathbf{\Sigma}\,\bm{B}(\cdot)$ is provided in Section \ref{sec:main results} and Lemma 2.3 in the supplemental material. 
For practical implementation, we estimate $\mathbf{\Sigma}$ and then substitute the estimator back into the expression; details of this procedure are given in Section \ref{sec:Testing the change point}.

The previous CUMSUM statistics mainly focus on the $\mathcal{L}_2$ norm (\cite{aue2018detecting}),  partially because the weak invariance principle is not hard to derive in Hilbert space, while Gaussian approximation in Banach space presents more challenge. For high-dimensional inference as discussed in \cite{li2024}, an {$\ell_2$-based test statistic is generally more powerful in detecting
weak dense signals compared to an $\ell_{\infty}$ one, while in the case of sparse signals, an $\ell_{\infty}$
type statistic appears better.} The same phenomenon also appears in function  space.

\par 
Once the testing procedure rejects   the null hypothesis of no abrupt change in the mean function, it is then of interest  to locate the break point. 
The  $\mathcal L_{\infty}$-based and $\mathcal L_{2}$-based break point estimators are defined accordingly,
\begin{align}\label{k_n2}
   & \widehat k_{n,2} = \mathop{\arg\max}
\limits_{\epsilon n\leq k\leq (1-\epsilon)n} \left\|  \bm B^\top(\cdot)\widehat{\mathbf V}^{-1}\left\{ \sum_{i=1}^{k}\frac{1}{N_i}\sum_{j=1}^{N_i}\bm B(X_{ij})Y_{ij}- \frac{k}{n}\sum_{i=1}^{n}\frac{1}{N_i}\sum_{j=1}^{N_i}\bm B(X_{ij})Y_{ij}
 \right\}\right\|_{\mathcal L_2}^2,\\
 & \widehat k_{n,\infty} = \mathop{\arg\max}
\limits_{\epsilon n\leq k\leq (1-\epsilon)n} \left\|  \bm B^\top(\cdot)\widehat{\mathbf V}^{-1}\left\{ \sum_{i=1}^{k}\frac{1}{N_i}\sum_{j=1}^{N_i}\bm B(X_{ij})Y_{ij}- \frac{k}{n}\sum_{i=1}^{n}\frac{1}{N_i}\sum_{j=1}^{N_i}\bm B(X_{ij})Y_{ij}
 \right\}\right\|_{\mathcal L_\infty}.\nonumber
\end{align}
When constructing the break point estimators, we avoid using the partial version of 
the inner product matrix $\widehat{\mathbf V}_t$, and instead use the global version of 
$\widehat{\mathbf V}$, which technically facilitates proving the 
convergence rate of the $\mathcal L_2$-based break point estimator $\widehat k_{n,2}$.  

\par
\begin{figure}[H] 
    \centering
    \includegraphics[scale =0.6]{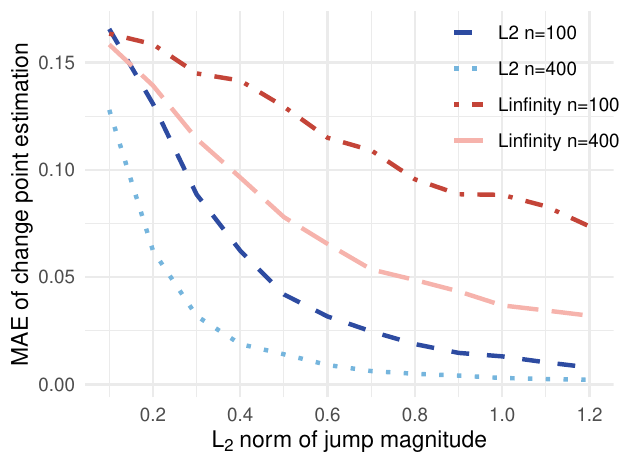}
    \includegraphics[scale =0.6]{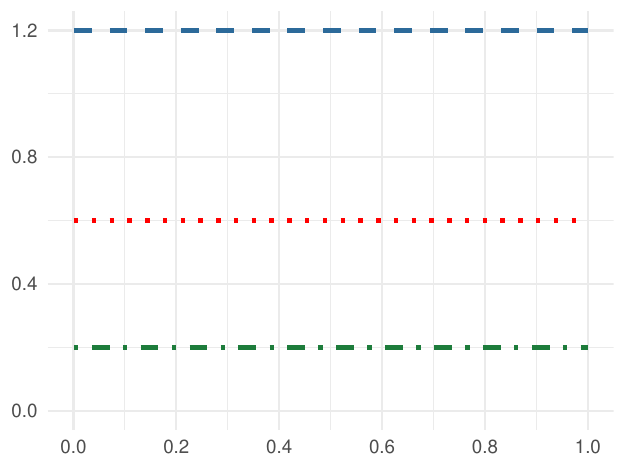}
    \includegraphics[scale =0.6]{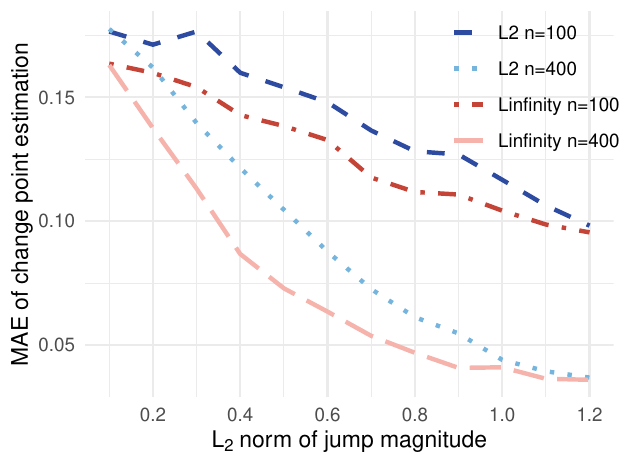}
    \includegraphics[scale =0.6]{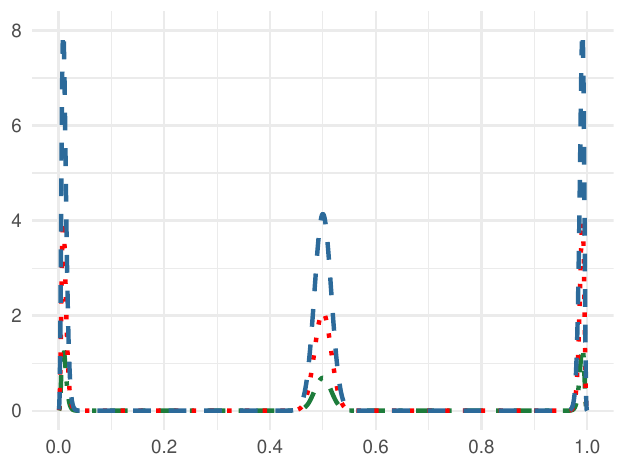}
    \caption{Examples of empirical performance of estimator $\widehat k_{n,\infty}$ and $\widehat k_{n,2}$ with MAE (left panel) with corresponding jump magnitude (right panel, $\mathcal L_2$ norm of jump magnitude, line for $1.2$, dotted for $0.6$, dashed for $0.2$).}
    \label{Fig:L2vsLinfty}
\end{figure}

  Although the convergence rate  of $\widehat k_{n,\infty}$ is  slower than  $\widehat k_{n,2}$ asymptotically in Theorem \ref{THM:consistency}, it has better empirical performance in special cases, as shown in Figure \ref{Fig:L2vsLinfty}. The jump magnitude are set as $\Delta(x) = a$ and $\Delta(x) = a\sqrt{\beta_{10,1000}(x) + \beta_{1000,1000}(x)  + \beta_{1000,10}(x) }/\sqrt{3}$, where $\beta_{\cdot,\star}(x)$ is the probability density function of $\beta$-distribution. Following the empirical results, one can find when $\|\Delta(x)\|_{\infty} = \|\Delta(x)\|_{\mathcal{L}_2}$ (constant functions), the performance of $\widehat{k}_{n,2}$ will be significantly better than $\widehat{k}_{n,\infty}$ due to the faster convergence rate. On the other hand, when $\|\Delta(x)\|_{\infty}$ much greater than $\|\Delta(x)\|_{\mathcal{L}_2}$  (functions with spark peak), the mean absolute error (MAE) of $\widehat{k}_{n,\infty}$ is  less than $\widehat{k}_{n,2}$. 


\par Moreover, given an  estimation of the break point, denoted by $\widehat{k}_n$, we aim to estimate  and conduct statistical inference for the jump function $\Delta(\cdot)$, as well as the mean functions before and after the break point. 
For the obtained break point estimation $\widehat{k}_n$, we estimate $\Delta(\cdot)$ as 
\begin{align}\label{Delta_hat}
    \widehat\Delta_n(\cdot) =  \widehat{m}_{\widehat{k}_n}(\cdot) -  \widehat{m}_{\widehat{k}_n+1}(\cdot),
\end{align}
where 
\begin{align} \label{m_kn}\widehat{m}_{\widehat{k}_n}(\cdot) &= \mathop{\arg\min}
\limits_{g(\cdot)\in \mathcal{S}_{J_n}^{p}}
\frac{1}{\widehat{k}_n}\sum_{i=1}^{\widehat{k}_n} \frac{1}{N_i}\sum_{j=1}^{N_i}  \left\{Y_{ij}-g
(X_{ij})\right\}^2,\\
\widehat{m}_{\widehat{k}_n+1}(\cdot) &= \mathop{\arg\min}
\limits_{g(\cdot)\in \mathcal{S}_{J_n}^{p}}
\frac{1}{n-\widehat{k}_n}\sum_{i=\widehat{k}_n+1}^{n} \frac{1}{N_i}\sum_{j=1}^{N_i}  \left\{Y_{ij}-g
(X_{ij})\right\}^2.\label{m_kn+1}
\end{align}

\section{Main results}\label{sec:main results}
\subsection{Asymptotic properties of structural break detectors}
For a real number $\nu\in(0,1]$ and integer $q\in\mathbb{N}$, $d\in\mathbb{N}_+$, write $\mathcal{H}_{q,\nu}\left([0,1]^d\right)$ as the space of $(q,\nu)$-H\"{o}lder continuous functions on $[0,1]^d$, that is,
	\begin{align*}
		\mathcal{H}_{q,\nu}\left([0,1]^d\right)=\left\{h:[0,1]^d\to \mathbb{R},~\left\|h\right\|_{q,\nu}=\sup_{t\neq s\in[0,1]^d ,\bm\alpha\in\mathbb N^d,\left\|\bm\alpha\right\|_1=q}\frac{\left|\partial^{\bm \alpha} h(t)-\partial^{\bm \alpha} h(s)\right|}{\left\|t-s\right\|_2^{\nu}}<\infty  \right\}.
	\end{align*}
\par To derive  asymptotic properties, some mild assumptions are introduced first. 
\begin{itemize}
    \item[(A1)] Assume that $m_i(\cdot)\in \mathcal{H}_{q,\nu}[0,1]$, $i = k_0, k_0+1$, for some integer $q>0$ and some positive real number $\nu\in(0,1]$. In the following, denote  $q^\ast=q+\nu$.
		\item[(A2)] The density function $f(\cdot)$ of $X$ is bounded above and away from zero, i.e., $c_{f}\leq \inf_{x\in[0,1]} f(x)  \leq \sup_{x\in[0,1]} f(x) \leq C_{f}$ for some positive constants $c_{f}$ and $C_{f}$.
  \item[(A3)] 
  The variance function $\sigma^2(\cdot)$  is uniformly bounded, and  $\E|\epsilon_i|^{r_1}<\infty$ for some integer $r_1\geq 3$.
  \item[(A4)] The long-run covariance function $G(\cdot,\cdot)\in\mathcal{H}_{0,\mu}[0,1]^2$ for some $\mu\in (0,1]$,   $ \inf_{x\in[0,1]}G(x,x)\geq c_{G}$ for some positive constant $c_{G}$, and $\sup_{x\in[0,1]}\E|\eta(x)|^{r_2}<\infty$ for some integer $r_2\geq 3$.
  \item[(A5)]
For some $\beta\geq 3$ and $r_2$ in (A4), the dependence measure $\delta_H(i,r_2)=\O\left(i^{-\beta}\right)$. 
\item[(A6)] Suppose that the spline order is $p\geq q^\ast$. Let $r=\min\{r_1,r_2\}$ and $a_n=\log^{1/2}n$. When $\E(N^{-1})J_n\gg 1$, $J_n^{-q^\ast-1/2}n^{1/2}\E^{-1/2}(N^{-1})a_n=\o(1)$, and  \begin{align*}
     J_n^{\frac{7r-6}{3r-2 }}\left\{\sum_{s=1}^{r}\E \left(N^{s-r}\right) J_n^{-s}\right\}^{2/r} n^{-\frac{r-2}{3r-2}} \E^{-1}(N^{-1})a_n^2\to 0.
\end{align*}

When   $\E(N^{-1})J_n\lesssim 1$, $J_n^{-q^\ast}n^{1/2}a_n=\o(1)$, and \begin{align*}
    J_n^{\frac{10r-8}{3r-2 }}\left\{\sum_{s=1}^{r}\E \left(N^{s-r}\right) J_n^{-s}\right\}^{2/r} n^{-\frac{r-2}{3r-2}}  a_n^2\to 0.
\end{align*}
 \item[(A6')] Let $a_n=1$ in (A6).
\end{itemize}
The smoothness of the mean functions and covariance function in Assumptions (A1) and (A4) are standard conditions in literature; see \cite{cao2012simultaneous} and  \cite{cai2024sparse} for instance. Assumption (A2) is common for random design, such as in \cite{10.1214/10-AOS813} and \cite{tonycai2011}.   Assumptions (A3) and (A4) impose moment conditions on the stochastic process $\eta_i(\cdot)$ and the measurement error $\epsilon_{ij}$, with similar conditions discussed in   \cite{ZW16} and  \cite{Zhang2021UnifiedPC}.
Since we do not impose a smoothness condition on $\sigma(\cdot)$, the partial measurement error framework in \cite{BHW24} can also be applied.
Assumption (A5) involves a polynomial rate of decay for the physical dependence measure, ensuring the weak dependence of the functional time series. Similar polynomial decaying rates can be also found in \cite{cui2023simultaneousinferencetimeseries} and \cite{zhou2023statistical}. 
The range of parameters specified in Assumptions (A1), (A3), and (A4) is detailed in Assumption (A6). 



Let $\mathcal{Z}_n(t,x)$ be a sequence of  Gaussian random fields  with mean zero and covariance function $ 
        {\mbox{Cov}}\left(\mathcal{Z}_n(t,x),\mathcal{Z}_n(t^\prime,x^\prime)\right) = \left(\min\{t,t^\prime\}-tt^\prime \right)\bm B^\top(x)\mathbf\Sigma\bm B(x^\prime)$.

\begin{theorem}\label{THM:general_H0}
   Suppose that the null hypothesis $H_0$ holds. \begin{itemize}
       \item[(a)] Under Assumptions (A1)-(A6),   as $n\to\infty$, 
 \begin{align}\label{Tn_H0}
         \sup_{z\in\mathbb R}\left|\P\left(T_n\leq z\right)-\P\left(\sup_{\epsilon\leq t\leq 1-\epsilon}\sup_{x\in[0,1]}   \frac{ \left|\mathcal{Z}_n(t,x) \right|}{\sqrt{ \bm B^\top(x)\mathbf\Sigma\bm B(x)}}\leq z\right)\right| \to 0.
         \end{align}
         \item[(b)] Under Assumptions (A1)-(A5) and (A6'),  as $n\to\infty$, 
         \begin{align}
              \label{Sn_H0}
        \sup_{z\in\mathbb R}\left|\P\left(S_n\leq z\right)-\P\left(\sup_{\epsilon\leq t\leq 1-\epsilon} \int_0^1 \frac{  \mathcal{Z}_n^2(t,x)  }{ \bm B^\top(x)\mathbf\Sigma\bm B(x)} dx\leq z\right)\right| \to 0.
    \end{align}
    \end{itemize}
\end{theorem}

Theorem \ref{THM:general_H0} describes  null distribution  of the test statistics $T_n$  in (\ref{EQ:Tn}) and $S_n$ in (\ref{EQ:Sn}), which is uniformly   approximated by  the distribution of certain functionals of a  sequence of zero-mean Gaussian fields. Since no restrictions are imposed on the relationship between $n$ and $N_i$ in Assumptions (A1)-(A6), Theorem \ref{THM:general_H0} contains scenarios with arbitrary sampling schemes  from sparse to dense.  An extreme case is $m_i(\cdot) \equiv m_i$. 
With a slight violation of the range of the number of knots in Assumption (A6),  the partial mean  (\ref{unweighted:mhat_t}) and global mean   (\ref{unweighted:mhat})  using the constant spline with $J_n+p = 1$ will degenerate to 
$\widehat{m}(t,\cdot) = [nt]^{-1}\sum_{i=1}^{[nt]}Y_{i1}, \widehat{m}(\cdot) = n^{-1}\sum_{i=1}^{n}Y_{i1}
$, and the term $\bm B^\top(x)\mathbf\Sigma\bm B(x^\prime)$  shrinks to a  constant scalar. In this case,   $T_n$  and $S_n$ weakly converge to the maximum absolute value and the maximum value of the square of Brownian bridges on $[ \epsilon, 1 - \epsilon ]$ respectively, as in the classical change point   theory in time series.

\par 
The covariance function of the approximating Gaussian processes can be decomposed into the sum of $\bm{B}^\top(x)\mathbf{\Sigma}_1\bm{B}(x')$ and $\bm{B}^\top(x)\mathbf{\Sigma}_2\bm{B}(x')$, reflecting on the definition of the matrix $\mathbf{\Sigma} = \mathbf{\Sigma}_1 + \mathbf{\Sigma}_2$. The  former represents the spline approximation of the long-run covariance function $G(x, x^\prime)$, interpreted as the influence of dependency of  $\left\{\xi_i(X_{ij})\right\}_{i,j=1}^{n,N_i}$  both within and between  curves $\{\xi_i(\cdot)\}_{i=1}^{n}$, while the latter reflects the influence of the variance of each observation $Y_{ij}$ on the nonparametric smoothing, with magnitude $J_n\E(N^{-1})$. Thus, as $n\to \infty$, the phase transition in our framework depends on the trade-off between  $\bm{B}^\top(x)\mathbf{\Sigma}_1\bm{B}(x')$ and $\bm{B}^\top(x)\mathbf{\Sigma}_2\bm{B}(x')$ by comparing the order of $J_n$ and $\E^{-1}(N^{-1})$, as discussed in Assumption (A6). 
However, phase transition results are typically expressed in terms of the relationship between the sample size and the sampling frequency in existing literature (\cite{ZW16}, \cite{Zhang2021UnifiedPC}), as this offers a more intuitive understanding. Following the idea in \cite{cai2024sparse}, we first specify the relationship among $J_n$, $n$, and $\E(N^{-1})$, and then translate the relationship between $J_n$ and $\E(N^{-1})$ into one between $n$ and $\E(N^{-1})$. 
Hence, the phase transitions can be concluded as the following corollary.



\begin{corollary}\label{COR:H0}
    Suppose that the null hypothesis $H_0$ holds, as $n\to\infty$, \begin{itemize}
        \item[(i)] Under Assumptions (A1)-(A6), when $\E^{-1}(N^{-1})\ll n^{1/(2q^\ast)}\log^{b-(2q^\ast+1)/q^\ast}n$ and $J_n\asymp \left\{n\E^{-1 }(N^{-1})\log^{2q^\ast b} n\right\}^{1/(2q^\ast+1)}$ for some $b>1/(2q^\ast)$, 
         \begin{align}
              \label{Tn_H0_sparse}
        \sup_{z\in\mathbb R}\left|\P\left(T_n\leq z\right)-\P\left(\sup_{\epsilon\leq t\leq 1-\epsilon}\sup_{x\in[0,1]}   \frac{ \left|\mathcal{Z}_{2n}(t,x) \right|}{\sqrt{ \bm B^\top(x)\mathbf\Sigma_2\bm B(x)}}\leq z\right)\right| \to 0.
    \end{align}
    \item[(ii)] Under Assumptions (A1)-(A6), when $n^{1/(2q^\ast)}\log^{b-(2q^\ast+1)/q^\ast}n\ll \E^{-1}(N^{-1})\ll   n^{1/(2q^\ast)}\log^{1/(2q^\ast)+2}n$ and  $J_n\asymp   n^{1/(2q^\ast)}\log^{b}n$ for some $b>1/(2q^\ast)$, 
         (\ref{Tn_H0}) holds.
    \item[(iii)] Under Assumptions (A1)-(A6), when $\E^{-1}(N^{-1})\gg n^{1/(2q^\ast)}\log^{1/(2q^\ast)+2}n$ and \\$n^{1/(2q^\ast)}\log^{1/(2q^\ast)}n\ll J_n\ll   \E^{-1 }(N^{-1})\log^{-2}n$, 
         \begin{align}
              \label{Tn_H0_dense}
        \sup_{z\in\mathbb R}\left|\P\left(T_n\leq z\right)-\P\left(\sup_{\epsilon\leq t\leq 1-\epsilon}\sup_{x\in[0,1]}   \frac{ \left|\mathcal{Z}_{1n}(t,x) \right|}{\sqrt{ \bm B^\top(x)\mathbf\Sigma_1\bm B(x)}}\leq z\right)\right| \to 0.
    \end{align}
         \item[(i')] Under Assumptions (A1)-(A5) and (A6'), when $\E^{-1}(N^{-1})\ll n^{1/(2q^\ast)}\log^{1/(2q^\ast)}n$ and $J_n\asymp \left\{n\E^{-1 }(N^{-1})\log n\right\}^{1/(2q^\ast+1)}$,
         \begin{align}
              \label{Sn_H0_sparse}
        \sup_{z\in\mathbb R}\left|\P\left(S_n\leq z\right)-\P\left(\sup_{\epsilon\leq t\leq 1-\epsilon} \int_0^1 \frac{  \mathcal{Z}_{2n}^2(t,x)  }{ \bm B^\top(x)\mathbf\Sigma_2\bm B(x)} dx\leq z\right)\right| \to 0.
    \end{align}
    \item[(ii')] Under Assumptions (A1)-(A5) and (A6'), when $\E^{-1}(N^{-1})\asymp  n^{1/(2q^\ast)}\log^{1/(2q^\ast)}n$ and  $J_n\asymp   n^{1/(2q^\ast)}\log^{1/(2q^\ast)}n$, 
         (\ref{Sn_H0}) holds.
    \item[(iii')] Under Assumptions (A1)-(A5) and (A6'), when $\E^{-1}(N^{-1})\gg n^{1/(2q^\ast)}\log^{1/(2q^\ast)}n$ and \\$n^{1/(2q^\ast)}\log^{1/(2q^\ast)}n\ll J_n\ll   \E^{-1 }(N^{-1})$, 
         \begin{align}
              \label{Sn_H0_dense}
        \sup_{z\in\mathbb R}\left|\P\left(S_n\leq z\right)-\P\left(\sup_{\epsilon\leq t\leq 1-\epsilon} \int_0^1 \frac{  \mathcal{Z}_{1n}^2(t,x)  }{ \bm B^\top(x)\mathbf\Sigma_1\bm B(x)} dx\leq z\right)\right| \to 0.
    \end{align}
    \end{itemize}
\end{corollary}
To better illustrate the intuition of  this result, we ignore logarithmic factors for simplicity and call $\sqrt{n/\bm B^\top(\cdot)\mathbf\Sigma\bm B(\cdot)}$ as the rescaling function of the proposed smoothed CUMSUM process. Overall, different leading terms in asymptotic covariance functions lead to different approximating Gaussian processes. Specifically,  when $\E^{-1}(N^{-1}) \ll n^{1/(2q^\ast)}$,   it is referred to as the sparse scenario, where $\bm B^\top(\cdot)\mathbf\Sigma_1\bm B(\cdot) \ll \bm B^\top(\cdot)\mathbf\Sigma_2\bm B(\cdot)$. Due to fewer observations along each curve, the influence of the variance of each $Y_{ij}$ dominates, and the magnitude of the rescaling function corresponds to the convergence rate of classical nonparametric regression. 
When $\E^{-1}(N^{-1}) \gg n^{1/(2q^\ast)}$,   it is referred to as the dense scenario, where $\bm B^\top(\cdot)\mathbf\Sigma_1\bm B(\cdot) \gg \bm B^\top(\cdot)\mathbf\Sigma_2\bm B(\cdot)$.
In this case, due to the sufficient number of observations on each curve,   the dependency between $\xi_i(X_{ij})$'s dominates, with the magnitude of the rescaling function being consistent with   the parametric convergence rate of $n^{1/2}$.
When $\E^{-1}(N^{-1}) \asymp n^{1/(2q^\ast)}$, it is referred to as the semi-dense scenario, where the two types of effects are mixed under this intermediate regime, similar to the findings in   \cite{berger2023dense}. The result of dense scenario  also coincides with the fact that  the test statistics in (\ref{EQ:Snfull}),  (\ref{EQ:dense}) and (\ref{EQ:Snsemidense}) all enjoy  parametric convergence rate of order $n^{1/2}$ when the sampling frequency is dense.

\par Next, we discuss the asymptotic properties under alternative hypothesis.  Consider local alternatives $H_{1n}: \Delta_n\neq 0, \,\,\left\|\Delta_n \right\|\to 0$, where $\left\|\cdot\right\|$ is $\sup$-norm or $\mathcal L_2$-norm. 
Define $\varrho_{\tau}(t) = t(1-\tau)$ if $t\leq \tau$, $\tau(1-t)$ if $t>\tau$ and $\tau_0=\lim_{n\to\infty}k_0/n$. 

\begin{theorem}\label{THM:general_H1}
    Suppose that the alternative hypothesis $H_{1}$ holds, as $n\to \infty$,
   \begin{itemize}
       \item[(a)] Under Assumptions (A1)-(A6),  when 
$\left\|\Delta_n\right\|_{\infty} \lesssim \max\left\{\sqrt{ n^{-1}{\log n} },  \sqrt{ {n^{-1}\E(N^{-1}) J_n\log n} }\right\}$,  
 \begin{align}\label{Tn_local_H1}
         \sup_{z\in\mathbb R}\left|\P\left(T_n\leq z\right)-\P\left(\sup_{\epsilon\leq t\leq 1-\epsilon}\sup_{x\in[0,1]}   \frac{ \left|\mathcal{Z}_n(t,x) + \sqrt{n}\varrho_{\tau_0}(t)\Delta_n(x)\right|}{\sqrt{ \bm B^\top(x)\mathbf\Sigma\bm B(x)}}\leq z\right)\right| \to 0.
         \end{align}
         For any $\alpha\in(0,1)$, when $\left\|\Delta_n\right\|_{\infty} \gtrsim \max\left\{\sqrt{ n^{-1}{\log n} },  \sqrt{ {n^{-1}\E(N^{-1}) J_n\log n} }\right\}$, $\P\left(T_n>Q_{T_n,1-\alpha}\right)>\alpha$; when $\left\|\Delta_n\right\|_{\infty} \gg \max\left\{\sqrt{ n^{-1}{\log n} },  \sqrt{ {n^{-1}\E(N^{-1}) J_n\log n} }\right\}$, $
            \P\left(T_n>Q_{T_n,1-\alpha}\right)\to 1$.  
         \item[(b)] Under Assumptions (A1)-(A5), (A6'), when  
$\left\|\Delta_n\right\|_{\mathcal L_2} \lesssim \max\left\{\sqrt{ n^{-1}  },  \sqrt{ {n^{-1}\E(N^{-1}) J_n } }\right\}$,    \begin{align}
              \label{Sn_local_H1}
        \sup_{z\in\mathbb R}\left|\P\left(S_n\leq z\right)-\P\left(\sup_{\epsilon\leq t\leq 1-\epsilon} \int_0^1 \frac{  \left|\mathcal{Z}_n(t,x)+\sqrt{n}\varrho_{\tau_0}(t)\Delta_n(x)\right|^2  }{ \bm B^\top(x)\mathbf\Sigma\bm B(x)} dx\leq z\right)\right|.
    \end{align}
    For any $\alpha\in(0,1)$, when $\left\|\Delta_n\right\|_{\mathcal L_2} \gtrsim \max\left\{\sqrt{ n^{-1}  },  \sqrt{ {n^{-1}\E(N^{-1}) J_n } }\right\}$, $\P\left(S_n>Q_{S_n,1-\alpha}\right)>\alpha$; when $\left\|\Delta_n\right\|_{\mathcal L_2} \gg \max\left\{\sqrt{ n^{-1}  },  \sqrt{ {n^{-1}\E(N^{-1}) J_n } }\right\}$, $
            \P\left(S_n>Q_{S_n,1-\alpha}\right)\to 1$.
   \end{itemize}
\end{theorem}

\par For any $\alpha\in(0,1)$,   the $\alpha$-quantiles $Q_{T_n, \alpha}$, $Q_{S_n, \alpha}$ are defined by \begin{align*}
     \P\left(\sup_{\epsilon\leq t\leq 1-\epsilon}\sup_{x\in[0,1]}\frac{\left|\mathcal Z_n(t,x)\right|}{\sqrt{\bm B^\top(x)\mathbf\Sigma\bm B(x)}}\leq Q_{T_n, \alpha}\right)=   
    \P\left(\sup_{\epsilon\leq t\leq 1-\epsilon}\int_0^1\frac{ \mathcal Z^2_n(t,x) }{ \bm B^\top(x)\mathbf\Sigma\bm B(x) }dx\leq Q_{S_n, \alpha}\right)= \alpha.
\end{align*} 
Theorem \ref{THM:general_H1}  provides meticulous power analysis of  $T_n$ and $S_n$ under the local alternatives. To avoid redundancy, we take $S_n$ as an example. Under dense or semi-dense scenarios that $\E(N^{-1})J_n\lesssim 1$, the test has nontrivial power when the jump magnitude exceeds an order of $n^{-1/2}$; while under sparse scenarios that $\E(N^{-1})J_n\gg 1$, the test has nontrivial power when the jump magnitude exceeds an order of $\E^{1/2}(N^{-1})J_n^{1/2}n^{-1/2}$.  
Therefore, for testing the  hypothesis in (\ref{EQ:hypothesis}),  Theorems (\ref{THM:general_H0}) and (\ref{THM:general_H1}) show that the decision rules $T_n \geq Q_{T_n, 1-\alpha}$ and $S_n \geq Q_{S_n, 1-\alpha}$ are consistent with significance level $\alpha$.

Furthermore, according to Theorem \ref{THM:general_H1}, it is interesting to note that  the $\mathcal L_2$-based test statistic $S_n$ theoretically  has a better power than the $\mathcal L_\infty$-based statistic $T_n$. However, our simulation experiments indicate that  the empirical performance of $T_n$ might surpass that of $S_n$ with finite sample sizes, especially when the jump function exhibits sharp peaks.

\subsection{Asymptotic properties of  break point estimators}
\par Next, we investigate the convergence rate of the proposed break point estimators.
\begin{theorem}\label{THM:consistency}
     Under Assumptions (A1)-(A5), $nJ_n^{-q^\ast}=\o(1+J_n\E(N^{-1}))$, $n^{-1/2}J_n\log(n)=\o(1)$, and $\left\|\Delta\right\|_{\mathcal L_2}>0$, 
    $\left| \widehat k_{n,2}-k_0 \right|=\O_p\left(  \max\left\{ 1,   J_n\E(N^{-1})   
 \right\}\right)$. 
 Under Assumptions (A1)-(A6) and $\left\|\Delta\right\|_{\mathcal L_\infty}>0$, $\left| \widehat k_{n,\infty}-k_0 \right|=\O_p\left( n^{1/2}\log^{1/2}(n) \max\left\{ 1,   J_n^{1/2}\E^{1/2}(N^{-1})   
 \right\}\right)$. 
\end{theorem}
Firstly, the convergence rate of $\widehat k_{n,2}$ in Theorem \ref{THM:consistency}  is optimal for the dense case where $\E(N^{-1})\lesssim J_n$, as the asymptotic distribution of the estimator for $\widehat k_{n,2}$ has been established in \cite{aue2018detecting} for fully observed trajectories, implying that $\widehat k_{n,2} - k_0$ weakly converges to some non-degenerate random variable. Secondly, leveraging the relationship among $J_n$, $n$, and $\E(N^{-1})$ in Corollary \ref{COR:H0}, the convergence rate of $\widehat k_{n,2}$ is faster by some  logarithmic order compared to  the detector developed in \cite{madrid2022change} from sparse to dense. Lastly, although the convergence rate of $\widehat{k}_{n,\infty}$ is sub-optimal, which is an artifact of the proof technique due to the absence of an inner product in the Banach space $C[0,1]$,  we empirically find that $\widehat k_{n,\infty}$  becomes more accurate than $\widehat k_{n,\infty}$ with finite sample sizes when the jump function shows sharp spikes, as shown in Figure \ref{Fig:L2vsLinfty}.

\subsection{Inference for the jump magnitude}

\par Given an estimation $\widehat k_n$ of break point, we then aim to conduct  simultaneous inference for the jump magnitude  $\Delta(\cdot)$, while    inference   for the mean functions before and after the true break point  $k_0$   follows a similar approach. 
Note that  $\widehat k_n$ can be specified by users. Our proposed estimator $\widehat k_{n,2}$   
in (\ref{k_n2})  is a default choice, but any consistent estimator that satisfies the following assumption can be used.
\begin{itemize}
    \item[(B1)] The estimator of the break point satisfies that $$\left|\widehat k_n -k_0 \right|=\o_p\left(\max\left\{  n^{1/2}\log^{-1/2}n, n^{1/2}J_n^{1/2}\E^{1/2}(N^{-1})\log^{-1/2}n 
 \right\}  \right).$$
\end{itemize}
\begin{theorem}\label{THM:jump_inference}
Suppose that the alternative hypothesis $H_1$ holds with $\tau_0\in[\epsilon,1-\epsilon]$ for some $\epsilon\in(0,1/2)$. 
Under Assumptions (A1)-(A6) and there exist Gaussian processes $\mathcal N_n(x)$'s with mean zero and covariance function  ${\mathrm{Cov}}\left(\mathcal N_n(x),\mathcal N_n(x^\prime)  \right)= \bm B^\top(x)\mathbf\Sigma \bm B(x^\prime)$, such that \begin{align*}
    \sup_{z\in\mathbb R}\left|\P\left( \left\|\frac{\sqrt{ n\{\tau_0(1-\tau_0)\}}\left\{\widehat\Delta_n(\cdot)-\Delta(\cdot)\right\} }{\sqrt{\bm B^\top(\cdot)\mathbf\Sigma \bm B(\cdot)}}\right\|_{\infty}\leq z\right)-\P\left(\left\| \frac{  \mathcal N_n(\cdot)}{\sqrt{\bm B^\top(\cdot)\mathbf\Sigma \bm B(\cdot)}}\right\|_{\infty}\leq z\right)\right|\to 0.
\end{align*}
\end{theorem}
Theorem \ref{THM:jump_inference} establishes Gaussian approximation for the jump estimator $\widehat{\Delta}_n(\cdot)$ in (\ref{Delta_hat}), valid for both sparse and dense functional time series data, which   shed lights on  constructing an asymptotically correct SCB with varying width for the jump magnitude.    Furthermore, one could easily establish the corresponding phase transition results, similar to those in Corollary \ref{COR:H0}, but omit them for brevity.

\section{Implementation}\label{sec:implement}

\subsection{Testing the break point}\label{sec:Testing the change point}

Given the estimation of the break point $\widehat k_n$, one computes  residuals $\widehat{U}_{ij} = \left\{Y_{ij}- \widehat{m}_{\widehat k_{n}}(X_{ij})\right\}\mathbf{1}_{\left\{i \leq \widehat k_{n}\right\}} + \left\{Y_{ij}- \widehat{m}_{\widehat k_{n}+1}(X_{ij})\right\}\mathbf{1}_{\left\{i>\widehat k_{n}\right\}}$. 
{\color{black} 
Define $\widehat{\mathbf\Sigma}_1 = \sum_{h=-L}^{L}\widehat{\mathbf\Sigma}_{1,h}$, and \begin{align*}
    \widehat {\boldsymbol \Sigma}_2=\widehat{\mathbf V}^{-1}\left[\frac{1}{n }\sum_{i=1}^{n}\frac{1}{N_i}\sum_{j=1}^{N_i}\bm B(X_{ij})\bm B^\top(X_{ij})\left\{\widehat U_{ij}^2+\sum_{h=-L,h\neq 0}^{L}\bm B^\top(X_{ij})\widehat{\mathbf\Sigma}_{1,h}\bm B(X_{ij})\right\}\right]\widehat{\mathbf V}^{-1},
\end{align*}
where  
\begin{equation*} 
    \begin{aligned}
        & \widehat {\boldsymbol \Sigma}_{1,h}=\widehat{\mathbf V}^{-1}\left\{\frac{1}{n }\sum_{i=1}^{n-h}\frac{1}{N_i N_{i+h}}\sum_{j=1}^{N_i}\sum_{j^\prime=1}^{N_{i+h}}\bm B(X_{ij})\bm B^\top (X_{i+h,j^\prime})\widehat U_{ij}\widehat U_{i+h,j^\prime}\right\}\widehat{\mathbf V}^{-1},\\
        &\widehat {\boldsymbol \Sigma}_{1,0}=\widehat{\mathbf V}^{-1}\left\{\frac{1}{n }\sum_{i=1}^{n}\frac{1}{N_i (N_{i}-1)}\sum_{j\neq j^\prime}^{N_i} \bm B(X_{ij})\bm B^\top (X_{ij^\prime})\widehat U_{ij}\widehat U_{ij^\prime}\right\}\widehat{\mathbf V}^{-1},
    \end{aligned}
\end{equation*}
and $L=L(n)$ diverges as $n$ goes to infinity. Based on the empirical guidelines from \cite{li2023detection} and \cite{HuStatisticalIF}, we consider values of $L = [n^{1/5}], \dots, [n^{1/5} \log \log n]$ in the numerical simulations. The numerical results were similar across these values, and we ultimately choose $L = [n^{1/5}]$ as the default. }
For the regular fixed design setting where $X_{ij}\equiv j/N$ and $N_i\equiv N$ for all $i=1,\ldots,n$, one could adopt the following simplified estimator of ${\mathbf\Sigma}_2$,   
\begin{align*}
    \widetilde{\mathbf\Sigma}_2 = \widehat{\mathbf V}^{-1}\left\{ \frac{1}{nN}\sum_{j=1}^{N}\bm B(j/N)\bm B^\top(j/N) \sum_{h=-L }^{L}\sum_{i=1}^{n}\widehat U_{ij}\widehat U_{i+h,j}\right\}\widehat{\mathbf V}^{-1}.
\end{align*}

To obtain the critical values of the proposed $\mathcal L_2$ test and  $\mathcal L_\infty$ test from sparse to dense, it suffices to obtain the quantile of the maximum value of the Gaussian process $\int_0^1 \{\bm B^\top(x)\mathbf\Sigma\bm B(x)\}^{-1}\mathcal{Z}_n^2(\cdot,x) dx$ and the Gaussian field  $\{\bm B^\top(\cdot)\mathbf\Sigma\bm B(\cdot)\}^{-1/2}|\mathcal{Z}_n(\star,\cdot)|$, on $[\epsilon,1-\epsilon]$ and $[\epsilon,1-\epsilon]\times[0,1]$ respectively, which is vital to generate  Gaussian process with mean zero, whose covariance structure mimics that of $\int_0^1 \{\bm B^\top(x)\mathbf\Sigma\bm B(x)\}^{-1}\mathcal{Z}_n^2(\cdot,x) dx$ and  $\{\bm B^\top(\cdot)\mathbf\Sigma\bm B(\cdot)\}^{-1/2}|\mathcal{Z}_n(\star,\cdot)|$ by the plug-in principle.
The following outlines an efficient computation method.  
Given the data $\{X_{ij},Y_{ij}\}_{i=1,j=1}^{n,N_i}$, one 
first computes the eigenvalues and eigenfunctions of covariance function $\{\bm B^\top(\star)\widehat{\mathbf\Sigma}\bm B(\star)\}^{-1/2} \{\bm B^\top(\star)\widehat{\mathbf\Sigma}\bm B(\cdot)\} \{\bm B^\top(\cdot)\widehat{\mathbf\Sigma}\bm B(\cdot)\}^{-1/2},$
denoted by $\{\widehat\vartheta_{kn},\widehat\varphi_{kn}(\cdot)\}_{k=1}^{N_s+p}$. Let $\widehat\kappa = \mathop{\arg\min}_{1\leq v\leq T}\left\{\sum_{k=1}^{v}\widehat{\vartheta}_{k}/\sum_{k=1}^{T}\widehat{\vartheta}_{k}>0.99\right\}$, where $\{\widehat\vartheta_k\}_{k=1}^{T}$ are the first $T$ estimated positive eigenvalues.   Then, one generates   i.i.d. Brownian bridges $\{\mathcal B_{b,k}(\cdot)\}_{b,k=1}^{B,\widehat\kappa}$,  and defines $\Xi_{b,n}(t)=\sum_{k=1}^{\widehat\kappa}\widehat\vartheta_k \mathcal B_{b,k}^2(t)$, $\Upsilon_{b,n}(t,x) = \sum_{k=1}^{\widehat\kappa}\widehat\vartheta_k^{1/2}\widehat{\varphi}_k(x) \mathcal B_{b,k}(t)$. 
Denote by $\widehat{Q}_{S_n, \alpha}$, $\widehat{Q}_{T_n, \alpha}$ the empirical upper-$\alpha$ quantile of the maximal absolute value for     $\{\Xi_{b,n}(\cdot)\}_{b=1}^{B}$ and $\{\Upsilon_{b,n}(\star, \cdot)\}_{b=1}^{B}$, respectively.  
Define 
\begin{align*}
   & \widehat T_n =  \sup_{\epsilon\leq t\leq 1-\epsilon}\left\|   \frac{  [nt]   \widehat{m}(t,\cdot)/ {\sqrt{n}}  -    {\sqrt{n}t} \widehat{m}(\cdot)  }{\sqrt{ \bm B^\top(\cdot)\widehat{\mathbf\Sigma}\bm B(\cdot)}}\right\|_{\infty}, 
   \quad \widehat S_n = \sup_{\epsilon\leq t\leq 1-\epsilon} \left\|   \frac{  [nt]   \widehat{m}(t,\cdot)/ {\sqrt{n}}  -    {\sqrt{n}t} \widehat{m}(\cdot)  }{\sqrt{ \bm B^\top(\cdot)\widehat{\mathbf\Sigma}\bm B(\cdot)}}\right\|_{\mathcal{L}_2}^2.
\end{align*}
Hence, an $\mathcal L_2$   ($\mathcal L_\infty$) test of significance level $\alpha$ rejects the null $H_{0}$ in (\ref{EQ:hypothesis}) if $\widehat S_n>\widehat{Q}_{S_n,\alpha}$    ($\widehat T_n>\widehat{Q}_{T_n,\alpha}$).

\subsection{Simultaneous confidence bands of the jump}
To construct the simultaneous confidence band of the jump magnitude $\Delta(\cdot)$, it suffices to obtain the quantile of the sup-norm of the Gaussian process $\{\bm B^\top(\cdot)\mathbf\Sigma\bm B(\cdot)\}^{-1/2}|\mathcal{N}_n(\cdot)|$.     
To obtain the estimated quantile, it is crucial to generate a Gaussian process with a mean of zero, whose covariance structure emulates that of 
$\{\bm B^\top(\cdot)\mathbf\Sigma\bm B(\cdot)\}^{-1/2}|\mathcal{N}_n(\cdot)|$ using the plug-in principle. 
One generates   i.i.d. $(J_n+p)$-dimensional Gaussian random vectors $\{\bm Z_{b,n}\}_{b=1}^{B}$ with mean zero and covariance matrix $\widehat{\mathbf \Sigma}$, independent of the data $\{X_{ij},Y_{ij}\}_{i=1,j=1}^{n,N_i}$,  and calculate the empirical upper-$\alpha$ quantile of the maximal absolute value for     $\{ \{\bm B^\top(\cdot)\widehat{\mathbf\Sigma}\bm B(\cdot)\}^{-1/2}\bm B^\top(\cdot)\bm Z_{b,n} \}_{b=1}^{B}$, denoted by $\widehat{Q}_{\Delta,\alpha}$. 
An asymptotic $100\left( 1-\alpha \right)\%$ SCB for $\Delta(x)$ over $x\in [0,1]$ is given by 
$\widehat{\Delta}_n( x)\pm n^{-1/2} \{\bm B^\top(\cdot)\widehat{\mathbf\Sigma}\bm B(\cdot)\}^{1/2} \widehat{Q}_{\Delta,\alpha},\,\, x\in
[0,1]$.

\section{Numerical simulation}\label{sec:simulation}
\subsection{General settings}

A total of four sampling schemes from sparse to dense are taken into consideration.  $N_i$ are i.i.d. from a discrete uniform distribution on the set, (1) $\{3,4,5,6\}$, (2) $\{\lfloor 2n^{1/5} \rfloor, \lfloor 2n^{1/5} \rfloor+1, \ldots, \lfloor 4n^{1/5} \rfloor\}$, (3) $\{\lfloor  n^{1/2} \rfloor, \lfloor n^{1/2} \rfloor+1, \ldots, \lfloor 2n^{1/2} \rfloor\}$, and (4) $\{\lfloor  n/8 \rfloor, \lfloor n/8 \rfloor+1, \ldots, \lfloor n/4 \rfloor\}$. We also evaluate 
three types of the jump function,   (i) $\Delta_{a}(x)\equiv a$, (ii) $\Delta_a(x)\equiv 4\sqrt{5}a(x-1/2)^2$, (iii) $\Delta_a(x) =   a*\sqrt{\beta_{10,1000}(x) + \beta_{1000,1000}(x)  + \beta_{1000,10}(x) }/\sqrt{3}$, in which $\beta_{a,b}(x)$ is the probability density function of $\beta$-distribution, each satisfying that $\left\|\Delta_a\right\|_{\mathcal L_2}=a$.  We consider $a=0,0.2,0.4,\ldots,1$,  where 
$a=0$ represents the null hypothesis, and 
$a>0$ represents the alternative hypothesis in (\ref{EQ:hypothesis}).     

 According to the Karhunen-Lo\`{e}ve  representation (  \cite{hsing2015theoretical}),
the data are generated from the following  model (\ref{model2}) by rewriting model (\ref{model1}) as \begin{align}\label{model2}
     Y_{ij} =m(X_{ij})+\sum_{k=1}^{\infty}\sqrt{\lambda_k}\xi_{ik}\psi_k(X_{ij})+\sigma(X_{ij})\epsilon_{ij}, \,\, 1\leq i\leq n, \,\,1\leq j\leq N_i.
 \end{align}
Here, the  eigenvalues $\lambda_{1}\geq \lambda_{2}\geq \cdots \geq 0$ satisfying $\sum_{k=1}^{\infty }\lambda_{k}<\infty $, and the 
corresponding eigenfunctions $\left\{ \psi_{k}\right\}_{k=1}^{\infty }$    form an orthonormal basis of $
\mathcal{L}^2\left[0,1\right]$. The random coefficients $\left\{ \xi_{ik}\right\}_{k=1}^{\infty }$, called FPC scores, are uncorrelated with mean $0$  and variance $1$. 
In each case, we set $m_1(x)=\cdots=m_{k_0}(x)=3/2\sin\{3\pi(x+ 1/2)\} + 2x^3$,    $m_{k_0+1}(x)=\cdots=m_{n}(x)=m_{k_0}(x)+\Delta_a(x)$ for $a=0,0.2,0.4,\ldots,1$, and  $\lambda_k=2^{1-k}$ for $k=1,\ldots,4$ and $\lambda_k=0$ for $k\geq 5$, $\psi_{2k-1}=\sqrt{2}\sin(2k\pi x)$ and $\psi_{2k}=\sqrt{2}\cos(2k\pi x)$ for $k\in\mathbb{Z}_+$. Besides, the FPC scores $\left\{\xi_{tk}\right%
\}_{t=1,k=1}^{n,4} $ are generated from the   MA(1) process, 
$\xi_{tk}=0.8\zeta_{tk} + 0.6\zeta_{t-1,k}$, 
where $\{\zeta_{tk}\}_{t=0,k=1}^{n,2}$ are i.i.d. random variables, and $X_{ij}$'s are independently generated from the uniform distribution $U(0,1)$. The measurement errors $\varepsilon_{ij}$'s are i.i.d. standard normal random variables, with the variance function    $\sigma^2(x) \equiv 1$.    We   consider different distributions of the FPC scores, including the standard normal distribution $N(0, 1)$, the uniform distribution $U(-\sqrt{3}, \sqrt{3})$, and the standardized Laplace distribution with density $f(x)=2^{-1/2} \exp(\sqrt{2}|x|)$. The number of sample size $n$ is set at   $200$  and $400$. The empirical replications $B$ in the quantile estimation and the Monte Carlo replications are both $500$.

 The cubic splines with $p=4$ are used in our numerical simulations. Besides, 
the number $J_n$ of  interior knots is an important smoothing parameter, and we recommend using a data-driven procedure to select $J_n \in [\min\{0.5(n\bar{N})^{1/9},0.5n^{1/8}\}, \max\{(n\bar{N})^{1/7},n^{1/6}\}]$ by minimizing the following BIC criterion:
\begin{align*} 
\mathrm{BIC}\left(J_{n}\right) &= \log\left(\frac{1}{n}\sum_{i = 1}^{n}\frac{1}{N_i}\sum_{j = 1}^{N_i}\left[\left\{Y_{ij} - \widehat{m}_{\widehat k_{n,2 },J_n}(X_{ij})\right\}^2\mathbf{1}\left\{i\leq \widehat k_{n,2 }\right\}\right.\right.\\&\left.\left.+\left\{Y_{ij} - \widehat{m}_{\widehat k_{n,2 }+1,J_n}(X_{ij})\right\}^2\mathbf{1}\left\{i> \widehat k_{n,2}\right\} \right] \right) + \frac{(J_n+p)\log(n)}{n},
\end{align*}
where $\widehat{k}_{n,2}$,     $\widehat{m}_{\widehat k_{n,2 },J_n}(\cdot)$  and $\widehat{m}_{\widehat k_{n,2 }+1,J_n}(\cdot)$ are defined in (\ref{k_n2}), 
 (\ref{m_kn}) and (\ref{m_kn+1}) respectively,  with the number $J_n$ of interior knots.

\subsection{Size and power examination}
From Tables \ref{tab:n=200}-\ref{tab:n=400}, we have the following findings. First,  the empirical size of our developed testing procedure is generally close to the predetermined significance level $\alpha=0.05$. 
Besides, as the strength of the jump signal (i.e., the value of $a$) increases, the empirical power of the proposed tests improves. Moreover, the empirical size and power of the tests are insensitive to the distribution of the stochastic process $\xi_i(\cdot)$.
\par 
Furthermore, regarding the power under the alternative hypothesis, the $\mathcal{L}_\infty$ statistic $T_n$ 
tends to exhibit superior performance when the sampling frequency is dense and the jump magnitudes are sharper, 
particularly for Type (iii) jump. In contrast, the $\mathcal{L}_2$ statistic $S_n$ 
is more effective under sparser sampling frequencies and flatter jump functions, especially for Type (i) jump.

\begin{table}[htbp]
  \centering
  \caption{Empirical size and power of the proposed $\mathcal L_2$ and $\mathcal L_{\infty}$ tests with  $n=200$.}
  \resizebox{0.9\textwidth}{0.7\textwidth}{
    \begin{tabular}{cccccccccccccc}
    \toprule
          &       & \multicolumn{6}{c}{$\mathcal L_2$}                        & \multicolumn{6}{c}{$\mathcal L_\infty$} \\
\cmidrule{3-14}    Jump  & FPC score/$a$ & 0     & 0.2   & 0.4   & 0.6   & 0.8   & 1     & 0     & 0.2   & 0.4   & 0.6   & 0.8   & 1 \\
       \midrule
          &       & \multicolumn{12}{c}{Setting (1)} \\
    \midrule
    \multirow{3}[1]{*}{Type (i)} & Normal & 0.044  & 0.156  & 0.586  & 0.964  & 1.000  & 1.000  & 0.056  & 0.212  & 0.498  & 0.826  & 0.972  & 0.998  \\
          & Uniform & 0.050  & 0.166  & 0.564  & 0.952  & 1.000  & 1.000  & 0.072  & 0.158  & 0.478  & 0.826  & 0.976  & 0.998  \\
          & Laplace & 0.076  & 0.144  & 0.606  & 0.960  & 1.000  & 1.000  & 0.086  & 0.194  & 0.474  & 0.806  & 0.986  & 1.000  \\
    \multirow{3}[0]{*}{Type (ii)} & Normal & 0.054  & 0.132  & 0.368  & 0.810  & 0.992  & 1.000  & 0.070  & 0.160  & 0.582  & 0.896  & 0.996  & 1.000  \\
          & Uniform & 0.056  & 0.112  & 0.358  & 0.806  & 0.990  & 1.000  & 0.070  & 0.188  & 0.556  & 0.892  & 0.996  & 1.000  \\
          & Laplace & 0.072  & 0.118  & 0.370  & 0.824  & 0.994  & 1.000  & 0.080  & 0.196  & 0.578  & 0.910  & 0.994  & 1.000  \\
    \multirow{3}[1]{*}{Type (iii)} & Normal & 0.054  & 0.086  & 0.120  & 0.256  & 0.482  & 0.652  & 0.074  & 0.234  & 0.344  & 0.516  & 0.694  & 0.790  \\
          & Uniform & 0.064  & 0.086  & 0.172  & 0.338  & 0.456  & 0.692  & 0.076  & 0.224  & 0.390  & 0.546  & 0.676  & 0.818  \\
          & Laplace & 0.064  & 0.088  & 0.160  & 0.306  & 0.500  & 0.674  & 0.076  & 0.230  & 0.370  & 0.548  & 0.736  & 0.812  \\
    \midrule
          &       & \multicolumn{12}{c}{Setting (2)} \\
    \midrule
    \multirow{3}[1]{*}{Type (i)} & Normal & 0.054  & 0.154  & 0.712  & 0.992  & 1.000  & 1.000  & 0.054  & 0.222  & 0.564  & 0.936  & 1.000  & 1.000  \\
          & Uniform & 0.060  & 0.166  & 0.688  & 0.998  & 1.000  & 1.000  & 0.066  & 0.208  & 0.588  & 0.908  & 0.998  & 1.000  \\
          & Laplace & 0.056  & 0.172  & 0.702  & 0.996  & 1.000  & 1.000  & 0.054  & 0.186  & 0.578  & 0.942  & 1.000  & 1.000  \\
    \multirow{3}[0]{*}{Type (ii)} & Normal & 0.060  & 0.126  & 0.464  & 0.930  & 1.000  & 1.000  & 0.070  & 0.224  & 0.718  & 0.964  & 0.998  & 1.000  \\
          & Uniform & 0.062  & 0.104  & 0.458  & 0.946  & 1.000  & 1.000  & 0.066  & 0.204  & 0.714  & 0.972  & 1.000  & 1.000  \\
          & Laplace & 0.064  & 0.092  & 0.470  & 0.948  & 1.000  & 1.000  & 0.050  & 0.214  & 0.728  & 0.980  & 1.000  & 1.000  \\
    \multirow{3}[1]{*}{Type (iii)} & Normal & 0.072  & 0.084  & 0.124  & 0.282  & 0.618  & 0.820  & 0.072  & 0.156  & 0.414  & 0.608  & 0.846  & 0.918  \\
          & Uniform & 0.058  & 0.088  & 0.152  & 0.300  & 0.556  & 0.816  & 0.066  & 0.148  & 0.390  & 0.632  & 0.816  & 0.942  \\
          & Laplace & 0.054  & 0.072  & 0.168  & 0.308  & 0.608  & 0.832  & 0.066  & 0.162  & 0.370  & 0.642  & 0.844  & 0.936  \\
    \midrule
          &       & \multicolumn{12}{c}{Setting (3)} \\
    \midrule
    \multirow{3}[1]{*}{Type (i)} & Normal & 0.076  & 0.164  & 0.910  & 1.000  & 1.000  & 1.000  & 0.074  & 0.246  & 0.764  & 0.992  & 1.000  & 1.000  \\
          & Uniform & 0.066  & 0.168  & 0.906  & 1.000  & 1.000  & 1.000  & 0.066  & 0.270  & 0.766  & 0.998  & 1.000  & 1.000  \\
          & Laplace & 0.074  & 0.170  & 0.886  & 1.000  & 1.000  & 1.000  & 0.058  & 0.248  & 0.792  & 0.996  & 1.000  & 1.000  \\
    \multirow{3}[0]{*}{Type (ii)} & Normal & 0.074  & 0.122  & 0.646  & 0.994  & 1.000  & 1.000  & 0.076  & 0.310  & 0.876  & 1.000  & 1.000  & 1.000  \\
          & Uniform & 0.072  & 0.146  & 0.654  & 1.000  & 1.000  & 1.000  & 0.070  & 0.320  & 0.892  & 0.998  & 1.000  & 1.000  \\
          & Laplace & 0.064  & 0.136  & 0.666  & 0.994  & 1.000  & 1.000  & 0.072  & 0.302  & 0.896  & 1.000  & 1.000  & 1.000  \\
    \multirow{3}[1]{*}{Type (iii)} & Normal & 0.062  & 0.068  & 0.154  & 0.436  & 0.836  & 0.978  & 0.064  & 0.196  & 0.590  & 0.900  & 0.996  & 1.000  \\
          & Uniform & 0.072  & 0.090  & 0.176  & 0.410  & 0.834  & 0.980  & 0.054  & 0.200  & 0.610  & 0.910  & 0.996  & 1.000  \\
          & Laplace & 0.070  & 0.076  & 0.142  & 0.470  & 0.818  & 0.972  & 0.056  & 0.202  & 0.602  & 0.930  & 0.990  & 1.000  \\
    \midrule
          &       & \multicolumn{12}{c}{Setting (4)} \\
    \midrule
    \multirow{3}[1]{*}{Type (i)} & Normal & 0.072  & 0.172  & 0.942  & 1.000  & 1.000  & 1.000  & 0.066  & 0.324  & 0.838  & 1.000  & 1.000  & 1.000  \\
          & Uniform & 0.064  & 0.172  & 0.942  & 1.000  & 1.000  & 1.000  & 0.076  & 0.288  & 0.824  & 1.000  & 1.000  & 1.000  \\
          & Laplace & 0.048  & 0.152  & 0.942  & 1.000  & 1.000  & 1.000  & 0.058  & 0.302  & 0.858  & 1.000  & 1.000  & 1.000  \\
    \multirow{3}[0]{*}{Type (ii)} & Normal & 0.076  & 0.186  & 0.758  & 1.000  & 1.000  & 1.000  & 0.070  & 0.382  & 0.908  & 1.000  & 1.000  & 1.000  \\
          & Uniform & 0.074  & 0.164  & 0.720  & 1.000  & 1.000  & 1.000  & 0.076  & 0.384  & 0.912  & 1.000  & 1.000  & 1.000  \\
          & Laplace & 0.074  & 0.150  & 0.736  & 1.000  & 1.000  & 1.000  & 0.064  & 0.396  & 0.940  & 1.000  & 1.000  & 1.000  \\
    \multirow{3}[1]{*}{Type (iii)} & Normal & 0.068  & 0.090  & 0.170  & 0.558  & 0.944  & 0.996  & 0.072  & 0.262  & 0.734  & 0.984  & 1.000  & 1.000  \\
          & Uniform & 0.070  & 0.094  & 0.178  & 0.534  & 0.934  & 0.996  & 0.064  & 0.228  & 0.752  & 0.978  & 1.000  & 1.000  \\
          & Laplace & 0.058  & 0.082  & 0.184  & 0.506  & 0.934  & 0.998  & 0.066  & 0.284  & 0.752  & 0.980  & 1.000  & 1.000  \\
    \bottomrule
    \end{tabular}%
    }
  \label{tab:n=200}%
\end{table}%

\begin{table}[htbp]
  \centering
  \caption{Empirical size and power of the proposed $\mathcal L_2$ and $\mathcal L_{\infty}$ tests with  $n=400$.}
    \resizebox{0.9\textwidth}{0.7\textwidth}{
    \begin{tabular}{cccccccccccccc}
    \toprule
          &       & \multicolumn{6}{c}{L2}                        & \multicolumn{6}{c}{Linfinity} \\
\cmidrule{3-14}    Jump  & FPC score/a & 0     & 0.2   & 0.4   & 0.6   & 0.8   & 1     & 0     & 0.2   & 0.4   & 0.6   & 0.8   & 1 \\
    \midrule
          &       & \multicolumn{12}{c}{Setting (1)} \\
    \midrule
    \multirow{3}[1]{*}{Type (i)} & Normal & 0.058  & 0.228  & 0.934  & 1.000  & 1.000  & 1.000  & 0.052  & 0.222  & 0.698  & 0.988  & 1.000  & 1.000  \\
          & Uniform & 0.050  & 0.296  & 0.944  & 1.000  & 1.000  & 1.000  & 0.046  & 0.220  & 0.682  & 0.976  & 1.000  & 1.000  \\
          & Laplace & 0.050  & 0.266  & 0.922  & 1.000  & 1.000  & 1.000  & 0.044  & 0.200  & 0.674  & 0.992  & 1.000  & 1.000  \\
    \multirow{3}[0]{*}{Type (ii)} & Normal & 0.044  & 0.148  & 0.750  & 0.996  & 1.000  & 1.000  & 0.052  & 0.264  & 0.844  & 0.992  & 1.000  & 1.000  \\
          & Uniform & 0.046  & 0.148  & 0.720  & 0.998  & 1.000  & 1.000  & 0.048  & 0.272  & 0.862  & 1.000  & 1.000  & 1.000  \\
          & Laplace & 0.044  & 0.124  & 0.730  & 0.994  & 1.000  & 1.000  & 0.058  & 0.272  & 0.848  & 0.998  & 1.000  & 1.000  \\
    \multirow{3}[0]{*}{Type (iii)} & Normal & 0.060  & 0.066  & 0.146  & 0.472  & 0.796  & 0.956  & 0.048  & 0.120  & 0.358  & 0.712  & 0.892  & 0.968  \\
          & Uniform & 0.054  & 0.076  & 0.176  & 0.484  & 0.758  & 0.932  & 0.046  & 0.144  & 0.384  & 0.670  & 0.866  & 0.962  \\
          & Laplace & 0.040  & 0.052  & 0.156  & 0.456  & 0.790  & 0.934  & 0.038  & 0.118  & 0.440  & 0.692  & 0.886  & 0.952  \\ \midrule
          &       & \multicolumn{12}{c}{Setting (2)} \\
    \midrule
    \multirow{3}[1]{*}{Type (i)} & Normal & 0.062  & 0.308  & 0.994  & 1.000  & 1.000  & 1.000  & 0.050  & 0.242  & 0.912  & 1.000  & 1.000  & 1.000  \\
          & Uniform & 0.056  & 0.258  & 1.000  & 1.000  & 1.000  & 1.000  & 0.048  & 0.264  & 0.922  & 1.000  & 1.000  & 1.000  \\
          & Laplace & 0.046  & 0.318  & 0.994  & 1.000  & 1.000  & 1.000  & 0.040  & 0.308  & 0.892  & 1.000  & 1.000  & 1.000  \\
    \multirow{3}[0]{*}{Type (ii)} & Normal & 0.066  & 0.176  & 0.926  & 1.000  & 1.000  & 1.000  & 0.044  & 0.330  & 0.970  & 1.000  & 1.000  & 1.000  \\
          & Uniform & 0.056  & 0.194  & 0.916  & 1.000  & 1.000  & 1.000  & 0.042  & 0.376  & 0.966  & 1.000  & 1.000  & 1.000  \\
          & Laplace & 0.052  & 0.176  & 0.922  & 1.000  & 1.000  & 1.000  & 0.050  & 0.348  & 0.958  & 1.000  & 1.000  & 1.000  \\
    \multirow{3}[1]{*}{Type (iii)} & Normal & 0.054  & 0.094  & 0.242  & 0.668  & 0.968  & 1.000  & 0.036  & 0.178  & 0.564  & 0.898  & 0.998  & 1.000  \\
          & Uniform & 0.058  & 0.064  & 0.244  & 0.662  & 0.986  & 0.998  & 0.038  & 0.146  & 0.592  & 0.908  & 0.998  & 1.000  \\
          & Laplace & 0.050  & 0.066  & 0.290  & 0.678  & 0.964  & 0.998  & 0.046  & 0.172  & 0.598  & 0.942  & 0.998  & 1.000  \\
    \midrule
          &       & \multicolumn{12}{c}{Setting (3)} \\
    \midrule
    \multirow{3}[1]{*}{Type (i)} & Normal & 0.066  & 0.440  & 1.000  & 1.000  & 1.000  & 1.000  & 0.056  & 0.378  & 0.996  & 1.000  & 1.000  & 1.000  \\
          & Uniform & 0.068  & 0.430  & 1.000  & 1.000  & 1.000  & 1.000  & 0.048  & 0.394  & 0.996  & 1.000  & 1.000  & 1.000  \\
          & Laplace & 0.070  & 0.466  & 1.000  & 1.000  & 1.000  & 1.000  & 0.068  & 0.400  & 0.996  & 1.000  & 1.000  & 1.000  \\
    \multirow{3}[0]{*}{Type (ii)} & Normal & 0.060  & 0.292  & 0.998  & 1.000  & 1.000  & 1.000  & 0.052  & 0.624  & 1.000  & 1.000  & 1.000  & 1.000  \\
          & Uniform & 0.054  & 0.318  & 0.998  & 1.000  & 1.000  & 1.000  & 0.046  & 0.620  & 0.998  & 1.000  & 1.000  & 1.000  \\
          & Laplace & 0.068  & 0.288  & 0.998  & 1.000  & 1.000  & 1.000  & 0.058  & 0.624  & 0.998  & 1.000  & 1.000  & 1.000  \\
    \multirow{3}[1]{*}{Type (iii)} & Normal & 0.050  & 0.098  & 0.428  & 0.970  & 1.000  & 1.000  & 0.044  & 0.338  & 0.926  & 1.000  & 1.000  & 1.000  \\
          & Uniform & 0.056  & 0.114  & 0.402  & 0.954  & 1.000  & 1.000  & 0.040  & 0.328  & 0.936  & 1.000  & 1.000  & 1.000  \\
          & Laplace & 0.052  & 0.110  & 0.380  & 0.956  & 1.000  & 1.000  & 0.054  & 0.298  & 0.942  & 0.998  & 1.000  & 1.000  \\
    \midrule
          &       & \multicolumn{12}{c}{Setting (4)} \\
    \midrule
    \multirow{3}[1]{*}{Type (i)} & Normal & 0.060  & 0.452  & 1.000  & 1.000  & 1.000  & 1.000  & 0.064  & 0.488  & 1.000  & 1.000  & 1.000  & 1.000  \\
          & Uniform & 0.064  & 0.468  & 1.000  & 1.000  & 1.000  & 1.000  & 0.052  & 0.498  & 1.000  & 1.000  & 1.000  & 1.000  \\
          & Laplace & 0.060  & 0.478  & 1.000  & 1.000  & 1.000  & 1.000  & 0.064  & 0.516  & 1.000  & 1.000  & 1.000  & 1.000  \\
    \multirow{3}[0]{*}{Type (ii)} & Normal & 0.058  & 0.322  & 1.000  & 1.000  & 1.000  & 1.000  & 0.066  & 0.724  & 1.000  & 1.000  & 1.000  & 1.000  \\
          & Uniform & 0.046  & 0.358  & 1.000  & 1.000  & 1.000  & 1.000  & 0.060  & 0.712  & 1.000  & 1.000  & 1.000  & 1.000  \\
          & Laplace & 0.062  & 0.366  & 1.000  & 1.000  & 1.000  & 1.000  & 0.064  & 0.726  & 1.000  & 1.000  & 1.000  & 1.000  \\
    \multirow{3}[1]{*}{Type (iii)} & Normal & 0.068  & 0.116  & 0.512  & 1.000  & 1.000  & 1.000  & 0.064  & 0.424  & 0.992  & 1.000  & 1.000  & 1.000  \\
          & Uniform & 0.060  & 0.112  & 0.522  & 0.998  & 1.000  & 1.000  & 0.068  & 0.438  & 0.992  & 1.000  & 1.000  & 1.000  \\
          & Laplace & 0.046  & 0.096  & 0.550  & 1.000  & 1.000  & 1.000  & 0.050  & 0.452  & 0.990  & 1.000  & 1.000  & 1.000  \\
    \bottomrule
    \end{tabular}%
    }
  \label{tab:n=400}%
\end{table}%

\subsection{SCB of jump magnitudes}

To assess the proposed inference procedure of jumps, we only consider two types (i) and (ii), which are generally smooth.  
Tables \ref{tab:SCB} show that in each case, the empirical
coverage rates of the proposed SCRs are close to the nominal confidence
level $1-\alpha$, a positive confirmation of the asymptotic theory.

\begin{table}[htbp]
  \centering
  \caption{Empirical coverage rates of  proposed $95\%$ SCB for the jump magnitude($a=1$).}
  \resizebox{!}{0.25\textwidth}{
    \begin{tabular}{cccccccc}
    \toprule
          &       & \multicolumn{3}{c}{Type (i)} & \multicolumn{3}{c}{Type (ii)} \\
\cmidrule{3-8}    $n$     & Setting/FPC score & Normal & Uniform & Laplace & Normal & Uniform & Laplace \\
    \midrule
    \multirow{4}[2]{*}{200} & Setting (1) & 0.936  & 0.956  & 0.946  & 0.938  & 0.948  & 0.942  \\
          & Setting (2) & 0.946  & 0.952  & 0.946  & 0.934  & 0.938  & 0.924  \\
          & Setting (3) & 0.952  & 0.930  & 0.944  & 0.938  & 0.950  & 0.940  \\
          & Setting (4) & 0.932  & 0.958  & 0.922  & 0.960  & 0.952  & 0.936  \\
    \midrule
    \multirow{4}[2]{*}{400} & Setting (1) & 0.954  & 0.964  & 0.956  & 0.968  & 0.954  & 0.962  \\
          & Setting (2) & 0.964  & 0.966  & 0.966  & 0.956  & 0.950  & 0.960  \\
          & Setting (3) & 0.950  & 0.958  & 0.952  & 0.950  & 0.962  & 0.952  \\
          & Setting (4) & 0.948  & 0.956  & 0.954  & 0.948  & 0.940  & 0.948  \\
    \bottomrule
    \end{tabular}%
    }
  \label{tab:SCB}%
\end{table}%

\section{Application}\label{sec:realdata}

\subsection{Electricity price-versus-demand  data}

We adopt the proposed methods to analyze the daily price-versus-demand curves of the German electricity power market in 2011. The data set records the hourly gross electricity demand in Germany  $\{X_{ij}\}_{i,j=1}^{n,N_i}$ and the average hourly   electricity prices 
$\{Y_{ij}\}_{i,j=1}^{n,N_i}$, and is available from Supplement B of \cite{10.1214/18-AOAS1230}.   Similar datasets from
different years have attracted much interest from statisticians and have been extensively studied in the  literature; see \cite{Liebl2013}, \cite{10.1214/18-AOAS1230}, \cite{10.1214/19-AOS1864} and \cite{xue2024change} for instance. 

After removing missing data points, we  construct the model in (\ref{model1}) by pairing electricity demand with electricity prices based on their timestamps to form a single data point, then treating the data points from each day as observations of a realization of a stochastic process. Then, the sample size and the number of observations in each trajectory  are given by $n=364$ and $N_i= 24$ for $i=1,\ldots,n$.  Firstly, we are
interested in   testing whether the mean function of the curve data undergoes an abrupt change. Our proposed $\mathcal L_2$ and $\mathcal L_{\infty}$ tests both strongly reject the null hypothesis, with p-values smaller than 0.001. Then, the we use the developed break point estimators $\widehat k_{n,2}$ and $\widehat k_{n,\infty}$  to locate the break point. Interestingly, both estimators identified the break point $n = 74$ as March 15, 2011. 
This can be explained as the result of the Fukushima nuclear leak in Japan on March 11, 2011. 
After the Fukushima incident, Germany permanently shut down $40\%$ of its nuclear power plants on March 15, leading to significant price fluctuations in the electricity market.  The reduction in nuclear power generation caused a tightening of electricity supply, resulting in a short-term increase in electricity prices, as similar results emphasized in \cite{10.1214/18-AOAS1230}.

\begin{figure}
    \centering
    \includegraphics[width=0.45\linewidth]{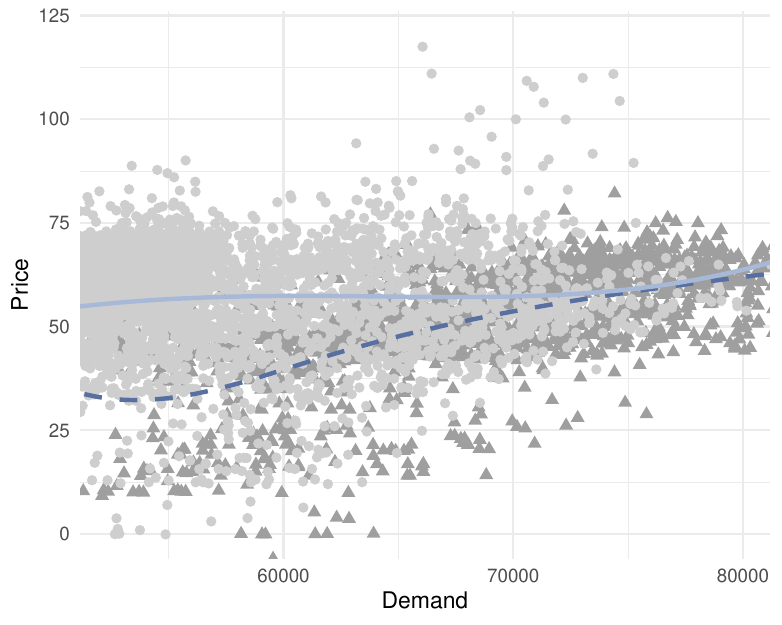}
    \includegraphics[width=0.45\linewidth]{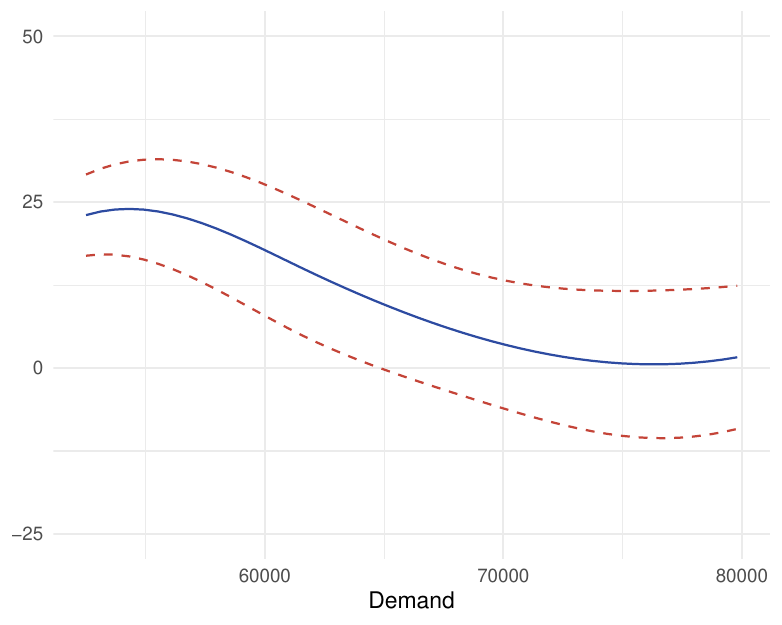}
    \caption{Left panel: The raw data before and after March 15, 2011 (before: triangle markers,  after: circle markers), and the corresponding estimation of the mean function (before: dashed line, after: solid line). Right panel: estimation of the jump function with the   $95\%$ SCB.}
    \label{fig:price-demand}
\end{figure}

The left panel in Figure \ref{fig:price-demand} shows the electricity price-demand curves before and after March 15, along with the estimated mean curves, further confirming the abrupt rise in Germany's electricity prices following the Japanese nuclear leak. 
While the right panel of Figure \ref{fig:price-demand} shows the estimation of the jump magnitude along with the corresponding $95\%$ SCB, indicating a significant increase in electricity prices at all levels except for high electricity demand.

\subsection{Temperature data}
We illustrate  our proposed method on densely observed functional times series by analyzing the temperature data from Sydney in Australia. 
The dataset we used records the daily maximum and minimum temperatures in Sydney from January 1, 1959, to December 31, 2008, and can be accessed on \url{http://www.bom.gov.au/climate/data/stations/}. break point  detection  problem for similar annual temperature curves in Australia has also been investigated in \cite{aue2018detecting}. 

\begin{figure}[H]
    \centering
    \includegraphics[width=0.4\linewidth]{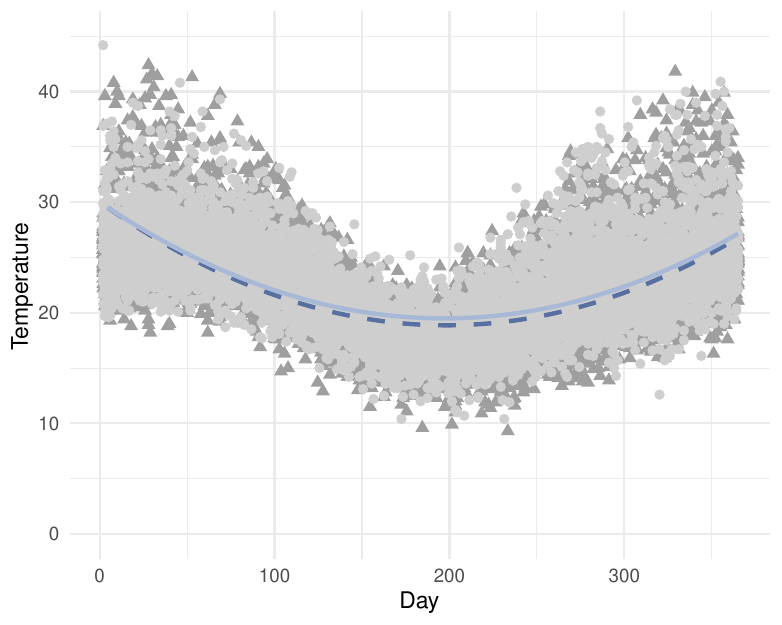}
    \includegraphics[width=0.4\linewidth]{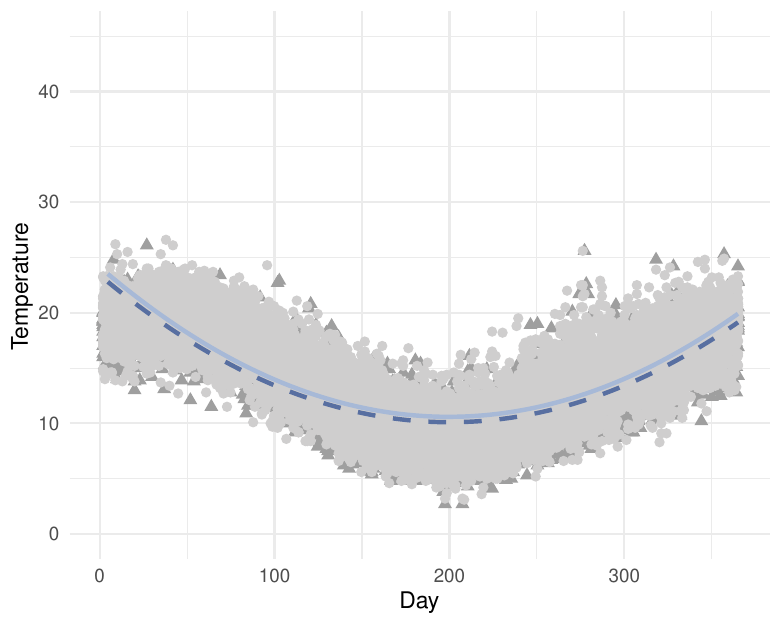}
    \includegraphics[width=0.4\linewidth]{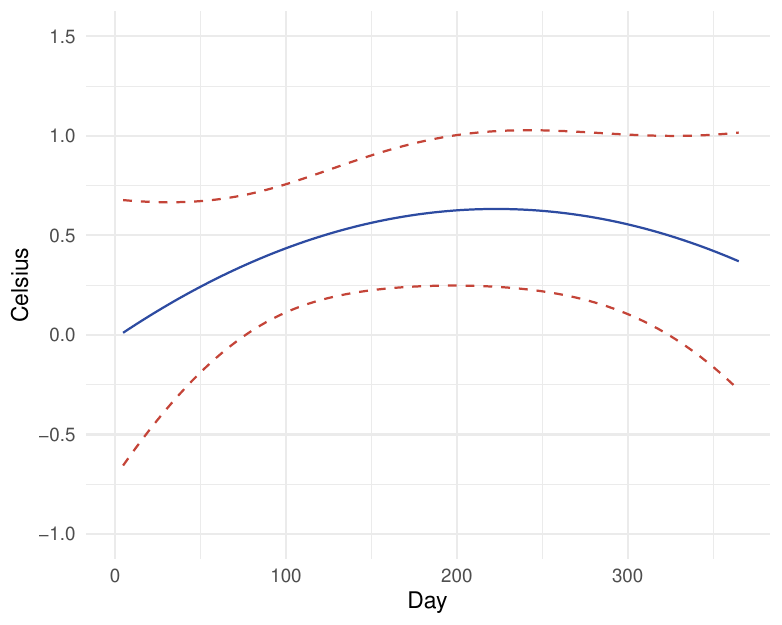}
    \includegraphics[width=0.4\linewidth]{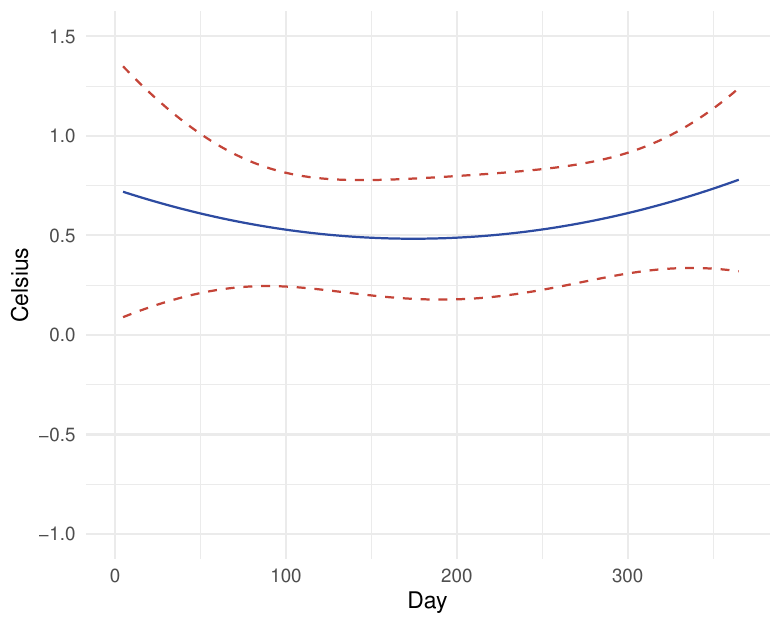}
    \caption{Top-left panel:   of the maximum temperature before and after 1988 in Sydney, and the corresponding estimation of the mean function.
     Top-right panel:   the minimum temperature before and after 1973 in Sydney, and the corresponding estimation of the mean function (before: triangle markers, dashed line;  after: circle markers, solid line).
     Bottom panels: estimation of the jump function  for the maximum temperature (left) and the minimum temperature (right) with the   $95\%$ SCB. }
    \label{fig:Sydney}
\end{figure}

After removing the missing data and the data points for February 29th in leap years, we can model the data as a functional time series in (\ref{model1}) with 
$n=50$ and 
$N_i\leq 365$ for $i=1,\ldots,n$.  
 The proposed break point testing procedures both reject the null hypothesis, suggesting   that a significant break point in Sydney's temperature climate occurred within  the half-century. 
 The location of the   year of temperature jump and the estimation  of the jump magnitude are summarized in Table \ref{table:Sydney} and Figure \ref{fig:Sydney}.
\begin{table}[htbp]
  \centering
  \caption{Summary of results for p-values and location of the break point in Sydney.}
    \resizebox{!}{0.1\textwidth}{
    \begin{tabular}{ccccc}
    \toprule
          & $\mathcal L_2$-test & $\mathcal L_\infty$-test & $\widehat k_{n,2}$    & $\widehat k_{n,\infty}$ \\
\cmidrule{2-5}    Minimum temperature & $<0.001$ & $<0.001$ & 1973  & 1972 \\
    Maximum temperature & $<0.001$ & $0.002$ & 1988  & 1986 \\
    \bottomrule
    \end{tabular}%
    }
  \label{table:Sydney}%
\end{table}%

Table \ref{table:Sydney} and Figure \ref{fig:Sydney} show that the daily minimum temperatures in nearly all seasons increased by $0.5$–$1$ Celsius around the early 1970s. This could be explained by breaks in the El Niño-Southern Oscillation (ENSO), which is often associated with higher temperatures and typically result in warmer temperatures and reduced rainfall. As \cite{trenberth19961990} and \cite{trenberth1997nino} noted, since the 1970s, breaks in the ENSO pattern began to emerge, with an increase in the frequency of El Niño events and a decrease in La Niña events. 
Additionally, the results of our method also indicate that the daily maximum temperatures also rose in the late 1980s, with the increase being particularly noticeable in winter, which coincides with the findings in \cite{aue2018detecting}.

\section{Concluding remark}
\par 
In this paper, we introduce a unified framework for detecting structural breaks in functional time series  from sparse to dense sampling designs. By constructing a smoothed CUMSUM process, we propose test statistics based on the $\mathcal{L}_2$- and $\mathcal{L}_\infty$-norms and derive their asymptotic properties by modern Gaussian approximation techniques. The corresponding break point estimators are obtained via the $\mathrm{argmax}$ functions of the proposed test statistics. Each of these two methods demonstrates either higher power or lower mean absolute error (MAE), respectively, under different types of jump magnitudes. We further develop a simultaneous confidence band for the jump magnitude as a global inference procedure that accommodates arbitrary sampling frequencies. Despite these contributions, several open problems remain. For instance, the inference of the break point itself and the sharp convergence rate for the $\mathcal{L}_\infty$ break point estimator warrant further investigation. In addition, our methodology can potentially be extended to other settings, such as functional spatial series or locally stationary time series, which provides  rich directions for future research.

\section*{Acknowledgements}
\par This research was supported by the National Natural Science Foundation of China awarded 12171269.

\bibliographystyle{apalike}
		\bibliography{ref}	

\begin{thebibliography}{}

\bibitem[Aue et~al., 2009]{aue2009estimation}
Aue, A., Gabrys, R., Horv{\'a}th, L., and Kokoszka, P. (2009).
\newblock Estimation of a change-point in the mean function of functional data.
\newblock {\em Journal of Multivariate Analysis}, 100(10):2254--2269.

\bibitem[Aue et~al., 2018]{aue2018detecting}
Aue, A., Rice, G., and S{\"o}nmez, O. (2018).
\newblock Detecting and dating structural breaks in functional data without
  dimension reduction.
\newblock {\em Journal of the Royal Statistical Society Series B: Statistical
  Methodology}, 80(3):509--529.

\bibitem[Baek et~al., 2024]{baek2024test}
Baek, C., Kokoszka, P., and Meng, X. (2024).
\newblock Test of change point versus long-range dependence in functional time
  series.
\newblock {\em Journal of Time Series Analysis}, 45(4):497--512.

\bibitem[Bai et~al., 2024]{BHW24}
Bai, L., Hu, Q., and Wu, W. (2024+).
\newblock Inference for structural changes in nonstationary functional time
  series with partial measurement error.
\newblock {\em Manuscript}.

\bibitem[Bastian et~al., 2023]{bastian2023multiple}
Bastian, P., Basu, R., and Dette, H. (2023).
\newblock Multiple change point detection in functional data with applications
  to biomechanical fatigue data.
\newblock {\em arXiv preprint arXiv:2312.11108}.

\bibitem[Berger et~al., 2023]{berger2023dense}
Berger, M., Hermann, P., and Holzmann, H. (2023).
\newblock From dense to sparse design: Optimal rates under the supremum norm
  for estimating the mean function in functional data analysis.
\newblock {\em arXiv preprint arXiv:2306.04550}.

\bibitem[Berkes et~al., 2009]{berkes2009detecting}
Berkes, I., Gabrys, R., Horv{\'a}th, L., and Kokoszka, P. (2009).
\newblock Detecting changes in the mean of functional observations.
\newblock {\em Journal of the Royal Statistical Society Series B: Statistical
  Methodology}, 71(5):927--946.

\bibitem[Bosq, 2000]{bosq2000linear}
Bosq, D. (2000).
\newblock {\em Linear processes in function spaces: theory and applications},
  volume 149.
\newblock Springer Science \& Business Media.

\bibitem[Cai and Hu, 2024a]{cai2024sparse}
Cai, L. and Hu, Q. (2024a).
\newblock From sparse to dense functional data: Phase transitions from a
  simultaneous inference perspective.
\newblock {\em arXiv preprint arXiv:2401.17646}.

\bibitem[Cai and Hu, 2024b]{caiHu}
Cai, L. and Hu, Q. (2024+b).
\newblock Simultaneous inference for distribution of functional principal
  scores.
\newblock {\em Statist. Sinica}.
\newblock DOI: 10.5705/ss.202023.0246.

\bibitem[Cai and Yuan, 2011]{tonycai2011}
Cai, T.~T. and Yuan, M. (2011).
\newblock {Optimal estimation of the mean function based on discretely sampled
  functional data: Phase transition}.
\newblock {\em The Annals of Statistics}, 39(5):2330--2355.

\bibitem[Cao et~al., 2012]{cao2012simultaneous}
Cao, G., Yang, L., and Todem, D. (2012).
\newblock Simultaneous inference for the mean function based on dense
  functional data.
\newblock {\em Journal of nonparametric statistics}, 24(2):359--377.

\bibitem[Chernozhukov et~al., 2015]{chernozhukov2015comparison}
Chernozhukov, V., Chetverikov, D., and Kato, K. (2015).
\newblock Comparison and anti-concentration bounds for maxima of gaussian
  random vectors.
\newblock {\em Probability Theory and Related Fields}, 162:47--70.

\bibitem[Cui and Zhou, 2023]{cui2023simultaneousinferencetimeseries}
Cui, Y. and Zhou, Z. (2023).
\newblock Simultaneous inference for time series functional linear regression.

\bibitem[De~Boor, 1978]{de1978practical}
De~Boor, C. (1978).
\newblock {\em A practical guide to splines}, volume~27.
\newblock springer-verlag New York.

\bibitem[Dette and Wu, 2024]{dette2021confidence}
Dette, H. and Wu, W. (2024+).
\newblock Confidence surfaces for the mean of locally stationary functional
  time series.
\newblock {\em Statistica sinica}.
\newblock DOI: 10.5705/ss.202023.0150.

\bibitem[Guo et~al., 2023]{guo2023sparse}
Guo, S., Li, D., Qiao, X., and Wang, Y. (2023).
\newblock From sparse to dense functional data in high dimensions: Revisiting
  phase transitions from a non-asymptotic perspective.
\newblock {\em arXiv preprint arXiv:2306.00476}.

\bibitem[Harris et~al., 2022]{harris2022scalable}
Harris, T., Li, B., and Tucker, J.~D. (2022).
\newblock Scalable multiple changepoint detection for functional data
  sequences.
\newblock {\em Environmetrics}, 33(2):e2710.

\bibitem[H{\"o}rmann and Kokoszka, 2010]{10.1214/09-AOS768}
H{\"o}rmann, S. and Kokoszka, P. (2010).
\newblock {Weakly dependent functional data}.
\newblock {\em The Annals of Statistics}, 38(3):1845--1884.

\bibitem[Horv{\'a}th and Kokoszka, 2012]{horvath2012inference}
Horv{\'a}th, L. and Kokoszka, P. (2012).
\newblock {\em Inference for functional data with applications}, volume 200.
\newblock Springer Science \& Business Media.

\bibitem[Horv{\'a}th et~al., 2013]{horvath2013estimation}
Horv{\'a}th, L., Kokoszka, P., and Reeder, R. (2013).
\newblock Estimation of the mean of functional time series and a two-sample
  problem.
\newblock {\em Journal of the Royal Statistical Society Series B: Statistical
  Methodology}, 75(1):103--122.

\bibitem[Hsing and Eubank, 2015]{hsing2015theoretical}
Hsing, T. and Eubank, R. (2015).
\newblock {\em Theoretical foundations of functional data analysis, with an
  introduction to linear operators}, volume 997.
\newblock John Wiley \& Sons.

\bibitem[Hu, 2024]{hu2024change}
Hu, Q. (2024).
\newblock Change point analysis of functional variance function with stationary
  error.
\newblock {\em Journal of Multivariate Analysis}, 202:105311.

\bibitem[Hu and Li, 2024]{HuStatisticalIF}
Hu, Q. and Li, J. (2024).
\newblock Statistical inference for mean function of longitudinal imaging data
  over complicated domains.
\newblock {\em Statistica Sinica}, 34:955--982.

\bibitem[Jiao et~al., 2023]{jiao2023enhanced}
Jiao, S., Chan, N.~H., and Yau, C.-Y. (2023).
\newblock Enhanced change-point detection in functional means.
\newblock {\em Statistica Sinica}.
\newblock DOI: 10.5705/ss.202022.0312.

\bibitem[Kneip and Liebl, 2020]{10.1214/19-AOS1864}
Kneip, A. and Liebl, D. (2020).
\newblock {On the optimal reconstruction of partially observed functional
  data}.
\newblock {\em The Annals of Statistics}, 48(3):1692--1717.

\bibitem[Li et~al., 2023]{li2023detection}
Li, D., Li, R., and Shang, H.~L. (2023).
\newblock Detection and estimation of structural breaks in high-dimensional
  functional time series.
\newblock {\em arXiv preprint arXiv:2304.07003}.

\bibitem[Li et~al., 2024]{li2024}
Li, J., Chen, L., Wang, W., and Wu, W.~B. (2024).
\newblock $\ell$2 inference for change points in high-dimensional time series
  via a two-way mosum.
\newblock {\em The Annals of Statistics}, 52(2):602--627.

\bibitem[Li and Hsing, 2010]{10.1214/10-AOS813}
Li, Y. and Hsing, T. (2010).
\newblock {Uniform convergence rates for nonparametric regression and principal
  component analysis in functional/longitudinal data}.
\newblock {\em The Annals of Statistics}, 38(6):3321--3351.

\bibitem[Liebl, 2013]{Liebl2013}
Liebl, D. (2013).
\newblock Modeling and forecasting electricity spot prices: A functional data
  perspective.
\newblock {\em The Annals of Applied Statistics}, 7(3):1562--1592.

\bibitem[Liebl, 2019]{10.1214/18-AOAS1230}
Liebl, D. (2019).
\newblock {Nonparametric testing for differences in electricity prices: The
  case of the Fukushima nuclear accident}.
\newblock {\em The Annals of Applied Statistics}, 13(2):1128--1146.

\bibitem[Madrid~Padilla et~al., 2022]{madrid2022change}
Madrid~Padilla, C.~M., Wang, D., Zhao, Z., and Yu, Y. (2022).
\newblock Change-point detection for sparse and dense functional data in
  general dimensions.
\newblock {\em Advances in Neural Information Processing Systems},
  35:37121--37133.

\bibitem[Mies and Steland, 2023]{mies2023sequential}
Mies, F. and Steland, A. (2023).
\newblock Sequential gaussian approximation for nonstationary time series in
  high dimensions.
\newblock {\em Bernoulli}, 29(4):3114--3140.

\bibitem[Rice and Zhang, 2022]{rice2022consistency}
Rice, G. and Zhang, C. (2022).
\newblock Consistency of binary segmentation for multiple change-point
  estimation with functional data.
\newblock {\em Statistics \& Probability Letters}, 180:109228.

\bibitem[Schumaker, 2007]{schumaker2007spline}
Schumaker, L. (2007).
\newblock {\em Spline functions: basic theory}.
\newblock Cambridge university press.

\bibitem[Sharghi Ghale-Joogh and Hosseini-Nasab, 2021]{sharghi2021mean}
Sharghi Ghale-Joogh, H. and Hosseini-Nasab, S. M.~E. (2021).
\newblock On mean derivative estimation of longitudinal and functional data:
  from sparse to dense.
\newblock {\em Statistical Papers}, 62(4):2047--2066.

\bibitem[Trenberth and Hoar, 1996]{trenberth19961990}
Trenberth, K.~E. and Hoar, T.~J. (1996).
\newblock The 1990-1995 el ni{\~n}o-southern oscillation event: Longest on
  record.
\newblock {\em Geophysical research letters}, 23(1):57--60.

\bibitem[Trenberth and Hoar, 1997]{trenberth1997nino}
Trenberth, K.~E. and Hoar, T.~J. (1997).
\newblock El ni{\~n}o and climate change.
\newblock {\em Geophysical Research Letters}, 24(23):3057--3060.

\bibitem[Wang et~al., 2023]{wang2023asynchronous}
Wang, M., Harris, T., and Li, B. (2023).
\newblock Asynchronous changepoint estimation for spatially correlated
  functional time series.
\newblock {\em Journal of Agricultural, Biological and Environmental
  Statistics}, 28(1):157--176.

\bibitem[Xue et~al., 2024]{xue2024change}
Xue, G., Xu, H., and Yu, Y. (2024).
\newblock Change point localisation and inference in fragmented functional
  data.
\newblock {\em arXiv preprint arXiv:2405.05730}.

\bibitem[Zhang and Li, 2022]{Zhang2021UnifiedPC}
Zhang, H. and Li, Y. (2022).
\newblock Unified principal component analysis for sparse and dense functional
  data under spatial dependency.
\newblock {\em Journal of Business \& Economic Statistics}, 40(4):1523--1537.

\bibitem[Zhang and Wang, 2016]{ZW16}
Zhang, X. and Wang, J.-l. (2016).
\newblock From sparse to dense functional data and beyond.
\newblock {\em The Annals of Statistics}, 44:2281--2321.

\bibitem[Zhou and Dette, 2023]{zhou2023statistical}
Zhou, Z. and Dette, H. (2023).
\newblock Statistical inference for high-dimensional panel functional time
  series.
\newblock {\em Journal of the Royal Statistical Society Series B: Statistical
  Methodology}, 85(2):523--549.

\end{thebibliography}
\end{document}